\documentstyle[12pt,epsfig]{article}
\newcommand{\beq}{\begin{equation}}
\newcommand{\eeq}{\end{equation}}
\def\bea{\begin{eqnarray}}
\def\eea{\end{eqnarray}}
\newcommand{\ba}{\begin{array}}
\newcommand{\ea}{\end{array}}

\begin{document}

\begin{titlepage}
\begin{flushright}
CERN-TH/97-111 \\
LPTHE-ORSAY/97-32\\
June 1997\\
\end{flushright}
\vspace{2cm}
\centerline{\Large A Wilson-Yukawa Model with undoubled chiral fermions in 2D}
\vspace{10pt}
\centerline{\Large\bf }
\vspace{24pt}
\centerline{\large P. Hern\'andez}
\centerline{\it Theory Division, CERN}
\centerline{\it CH-1211 Geneva 23, Switzerland}
\vspace{2pt}
\centerline{\large Ph. Boucaud}
\centerline{\it LPTHE, Universit\'e de Paris XI}
\centerline{\it 91405 Orsay Cedex, France.}

\vspace{4cm}
\begin{abstract}
We consider the fermion mass spectrum in the strong coupling vortex phase 
(VXS) of a lattice fermion-scalar model with a global $U(1)_L\times U(1)_R$, 
in 
two dimensions, in the context of a recently proposed two-cutoff lattice 
formulation. The fermion doublers are easily made massive 
by a strong Wilson-Yukawa coupling, but in contrast with the standard
formulation of these models, in which the light fermion spectrum was found 
to be vector-like, we find massless fermions with chiral quantum numbers at 
finite lattice spacing. When the global symmetry is gauged, this model 
is expected to give rise to a lattice chiral gauge theory.
 
\end{abstract}
%\vspace{5cm}
%CERN-TH/97-111\\
%LPTHE-ORSAY/97-32\\
%June 1997
\end{titlepage}
%\

\renewcommand{\thefootnote}{\arabic{footnote}}
\setcounter{footnote}{0}

\section{Introduction}
Wilson-Yukawa models were extensively studied some years ago as a possible 
lattice formulation of chiral gauge theories
\cite{wy}. The idea was to 
decouple the lattice doublers from the light spectrum by introducing 
the so-called 
Wilson-Yukawa term, which contains a charged scalar field in a way that it is 
manifestly gauge invariant under a chiral gauge transformation. With the 
addition of this ``Higgs'' particle, it was expected that doublers would 
decouple, while chiral fermions would remain in the light spectrum. 
This expectation was based on the numerical finding of a strong Yukawa-coupling regime, both in the symmetric (PMS) and broken (FMS) phases, where 
the fermion masses
are not proportional to the vacuum expectation value of the scalar field,
 as in the perturbative regime, and doublers can get
masses of the order of the cutoff. 
However, for the same reason, also the light fermions are massive in this 
phase (although 
their mass could be tuned to zero by tuning the bare mass).
More problematic was the fact that the light fermions were found 
to have only vector-like couplings to the gauge fields. 
The physical picture that emerged 
from these results was that in the strong coupling phase,  
composites of the light fermions and the 
scalar particles were formed, generating effective mirror states and 
rendering the theory vector-like. This situation was found already in the 
quenched approximation. 

It is important to realize that 
the failure of Wilson-Yukawa models is also very relevant for all
those lattice proposals of chiral gauge theories that explicitly break 
the gauge symmetry. The reason is that, once gauge invariance has been 
broken for a compact group, the gauge non-invariant theory is exactly 
equivalent to a gauge-invariant theory with new charged scalar degrees of 
freedom \cite{omega}. The scalar degrees of freedom are nothing more  than the pure gauge 
transformations. The effective action of these scalars differs in general from
one proposal to another, but they are 
coupled as long as gauge invariance is broken in the original formulation 
 and so they must be dealt with. In particular, the simplest formulation of 
chiral gauge theories, which introduces a standard Wilson term to decouple doublers, breaks gauge invariance and is in one-to-one 
correspondence to the 
Wilson-Yukawa models considered in \cite{wy}. Equivalently all 
proposals that break gauge invariance
must have a corresponding Wilson-Yukawa picture. It is important 
to understand to what extent the failures of these models in providing 
a regularization for chiral gauge theories are circumvented by other 
proposals. 

Recently a  new method has been proposed to regulate chiral gauge theories 
on the lattice \cite{us1}. In this approach, the fermions 
live on a lattice of spacing $f$ and are 
coupled to gauge link variables that are constructed by an appropriate 
smooth and gauge-invariant interpolation  of gauge configurations 
\cite{us2}
that are generated on a coarser lattice of spacing $b$. This construction
 corresponds to a two-cutoff lattice regulator (TC), where fermion momenta 
are cut off at a much larger scale $1/f$ than the gauge boson momenta $1/b$. 
Doublers are decoupled by introducing a naive Wilson term as in the 
Roma approach \cite{roma}. However, since the 
chirally breaking effects due to the Wilson term are relevant only at 
scales of the order of the fermion cutoff,  gauge boson momenta are not large 
enough to prove these interactions, in contrast with one-cutoff (OC) formulations. As a result, due to the separation of 
the cutoff scales, no fine-tuning is necessary  
to recover an approximate chiral gauge symmetry, as long as the ratio $f/b$ 
is small enough. 
In other words, gauge invariance is broken but there is a small 
parameter, $f/b$, which controls 
the strength of this breaking. 
As we will see, there are good arguments to
believe that a ``small'' breaking of gauge invariance does not have a dramatic 
effect at large distances \cite{nielsen}. The two-cutoff construction 
is the first proposal 
that has a control over the strength of this breaking.

In this paper we want to consider the Wilson-Yukawa picture of the 
TC formulation of \cite{us1} and show evidence that in fact a chiral phase 
exists, in contrast to what was found in OC formulations. 
In order to address these issues numerically, we have considered the 
simplest relevant model, which is a $U(1)$ compact theory in two dimensions, 
in the quenched approximation. The model contains one left-handed fermion
coupled to $U(1)$ gauge fields \footnote{When the quenched approximation is 
relaxed more species are needed to cancel gauge anomalies.}. 
In order to get rid of doublers, a right-handed
fermion is introduced, which only couples through the Wilson term.
In the global limit ($g\rightarrow 0$, i.e. only the pure gauge transformations are coupled) 
the Wilson-Yukawa picture is a scalar-fermion model with
 an exact global symmetry $U(1)_L \times U(1)_R$.  
This model has been previously considered in a standard OC formulation
in \cite{smit1}, where it was found that the fermion spectrum was either
doubled (in the weak Yukawa phase) or vector-like (in the strong Yukawa phase).
We will show that in the TC formulation a truly chiral phase exists.

The paper is organized as follows.
In section 2, we review the relation between gauge-non-invariant formulations
of chiral gauge theories and models of the Wilson-Yukawa type. 
In section 3, we present the two-cutoff lattice formulation
 of the two-dimensional
model under study and its equivalent Wilson-Yukawa picture. 
In section 4, we discuss  the expected phases of this model and why
a phase with massless chiral fermions should exist.
In section 5, we present our numerical results on the fermion spectrum,
which support the existence of a chiral phase. In section 6 we briefly 
discuss the effects of unquenching and conclude in section 7.

\section{Wilson-Yukawa picture of $\chi GTh$} 

As explained in the introduction, all proposals to regulate chiral 
gauge theories which break the gauge symmetry  
are equivalent to a model with an exact gauge invariance and additional 
charged scalars \cite{omega}. This is simple to see. Let  us consider a lattice action that contains gauge-breaking interactions,
\begin{eqnarray}
S[ U, \Psi] = S_{g.i.}[U, \Psi] + \delta S_{n.g.i.}[U,\Psi]
\end{eqnarray}
The path integral is
\begin{eqnarray}
Z = \int {\cal D}U_{\mu} \int d\Psi d\bar{\Psi} \;\; e^{-S_{g.i}[U,\Psi]-\delta S_{n.g.i.}[U,\Psi]}.
\label{ni}
\end{eqnarray}
Since the group is compact we can multiply $Z$ by the volume 
of the group $\int {\cal D}\Omega$, which is an irrelevant constant factor:
\begin{eqnarray}
Z = \int {\cal D}\Omega \int {\cal D} U_{\mu} \int d\Psi d\bar{\Psi}\;\; e^{S_{g.i}[U,\Psi]+\delta S_{n.g.i.}[U,\Psi]}.
\end{eqnarray}
Performing a change of variables, from $U_{\mu}$ to $U^{\Omega}_{\mu} \equiv 
\Omega(x) U_{\mu}(x) \Omega^\dagger(x+\hat{\mu})$ 
(i.e. the gauge-transformed variables under 
the lattice gauge transformation $\Omega$), and from $\Psi$ to $\Psi^{\Omega(x)} = \Omega(x) \Psi(x)$ (only for the charged fermions), and using the invariance of the
measure and $S_{g.i.}$ under a gauge transformation\footnote{We are on a lattice, so even if the fermions are chiral the fermion lattice measure is invariant
under the unitary transformation under $\Omega$.}, we get
\begin{eqnarray}
Z = \int {\cal D}\Omega \int {\cal D}U_{\mu} \int d\Psi d\bar{\Psi} \;\;e^{S_{g.i}[U,\Psi]+\delta S_{n.g.i.}[U^\Omega,\Psi^\Omega]}.
\label{gi}
\end{eqnarray}
It is easy to see that (\ref{gi}) is gauge-invariant under a new gauge 
symmetry under which the field $\Omega$ transforms
\begin{eqnarray}
U_{\mu}(x) &\rightarrow & \Phi(x) U_{\mu}(x) \Phi^\dagger(x+\hat{\mu}) \;\;\;\; \Omega(x)  \rightarrow  \Omega(x) \Phi(x)^\dagger \nonumber\\
\Psi^{(c)} &\rightarrow & \Phi(x) \Psi^{(c)}(x)  \;\;\;\;\;\;\;\; \Psi^{(n)} \rightarrow \Psi^{(n)}(x),
\label{gtg}
\end{eqnarray}
where $c,n$ are charged and neutral fermion fields respectively. This is 
because $\Psi^\Omega$ and $U_{\mu}^\Omega$ are invariant under this 
transformation.
This theory is then a gauge theory with extra charged scalars in a non-linear
realization ($\Omega$ is unitary). These scalars
are the pure gauge transformations, which couple through the non-invariant terms in the original action. 
It is clear that in any successful proposal for regulating chiral gauge 
theories, 
these scalars must decouple from the light physical spectrum. In the 
case of a spontaneously broken gauge theory 
these degrees of freedom remain, since they become the longitudinal 
gauge bosons. This is however not our objective here, since we aim at
regulating a symmetric chiral gauge theory. The interesting case of 
spontaneous symmetry breaking will be considered elsewhere.

Foester, Nielsen and Ninomiya (FNN) gave in \cite{nielsen} a good argument 
why a gauge
theory should be recovered at long distances in theories like (\ref{ni}), provided that the strength of the interactions in $\delta S_{n.g.i.}$ is ``small''. 
Their argument is quite simple. Let us suppose that the characteristic coupling
of the  non-invariant terms is arbitrarily small at the cutoff scale. 
Then in the gauge-invariant picture 
 of the theory (\ref{gi}), this implies that the $\Omega$ fields are 
very weakly coupled. Consequently, the free energy will be maximized when the 
$\Omega$ variables are decorrelated at distances of the order of the cutoff (i.e.
the lattice spacing)  or, 
in other words, these degrees of freedom are effectively very massive. 
Then it makes sense to integrate them out in order to obtain an effective 
theory at low energies. At distances much larger than the cutoff, but 
still smaller than 
the correlation length of the gauge-invariant degrees of freedom, it can 
be argued that 
the effective action should be local (after all it comes from the 
integration of heavy particles) and exactly gauge-invariant (it is clear
from eq. (\ref{gi}) that if we perform the integration over $\Omega$ what 
remains is an exactly gauge-invariant theory due to the exact symmetry
 (\ref{gtg})). In this situation, the only effect of the $\Omega$ fields
is a renormalization of the gauge-invariant coupling in $S_{g.i.}$. 

The difficult part in realizing the FNN mechanism is to satisfy
the requirement that the interactions of the $\Omega$ fields are ``small''.
The typical situation is that the $\Omega$ fields are coupled through a 
higher-dimensional operator, like a Wilson term, 
which is suppressed by the fermion cutoff.  
However, an operator of this kind does not satisfy the requirement. 
This can be seen already at the perturbative level. Although the Wilson term is an irrelevant
operator, which vanishes naively in the continuum limit, it  
appears in ultraviolet divergent 
loops, inducing corrections to the scalar action of $O(1)$ at all orders. 
 In contrast,
if the boson cutoff ($\Omega$ and $U$ fields) is much smaller than the fermion cutoff,
it can be shown that the Wilson term is really an irrelevant coupling at 
long distances, because the gauge breaking effects induced by 
loop corrections 
are suppressed by powers of the ratio of the boson cutoff to the 
fermion one, to all orders. For details on the power counting arguments 
specific to 
the lattice TC formulation,
 we refer the reader to \cite{us1} and \cite{us2}.
The FNN conditions are then 
satisfied by the TC formulation if the ratio of these cutoffs is small 
enough, and 
we expect that the mechanism is realized and an effective gauge invariance
is recovered at large distances.  

\section{The two-cutoff lattice formulation.}

In \cite{us1} and \cite{us2}, we presented a two-cutoff lattice construction of 
chiral gauge theories. A different proposal in the same direction was given
in \cite{frolov}. The advantage of using two-cutoffs as we have explained 
is precisely 
making the violations of gauge invariance small at large distances. With 
one cutoff a non-perturbative tunning of counterterms  would
be needed to achieve this \cite{roma}. 

The way in which the cutoff separation is implemented is by the use of two
lattices. The gauge degrees of freedom are the Wilson link variables of 
an euclidean lattice of spacing $b$, ${\cal L}_b$. We will call $s$ the
sites of the $b$-lattice. The gauge action is the standard one, 
\begin{eqnarray}
S_{g} = \frac{2}{g^2} \sum_s \sum_{\mu<\nu} [I - \frac{1}{2}(U_{\mu\nu}+ U_{\mu\nu}^\dagger)] 
\nonumber\\
U_{\mu\nu} \equiv U_\mu(s) 
U_\nu(s+\hat{\mu})
U_\mu^{-1}(s+\hat{\nu}) U_\nu^{-1}(s).
\label{gauact}
\end{eqnarray}
Fermions on the other hand 
live on the sites of a finer lattice ${\cal L}_f$ (some integer subdivision of ${\cal L}_b$). We refer to the $f$-lattice sites as $x$. 
In order to decouple the unavoidable doublers, a Wilson term is included in 
the fermionic action. For each charged chiral fermion, a singlet of the opposite 
chirality is needed. 
The fermions are coupled to the gauge fields through a standard
lattice gauge-fermion coupling on the $f$-lattice. The $f$-link variables
are obtained through a careful interpolation of the 
real dynamical fields, i.e. $U_{\mu}(s)$. As long
as the interpolation is smooth, it is clear that the separation of scales is
achieved in this construction, since the high momentum modes of the gauge fields
are cut off at the scale $b^{-1}$, while the fermions can have momenta of 
$O(f^{-1})$. 
 The 
lattice action for a charged left-handed fermion field is then given by
\begin{eqnarray}
S_{f} = \frac{1}{2} \sum_{x,\mu} \bar{\Psi}(x) \gamma_\mu 
[P_R + u_\mu(x) P_L]
\Psi(x+\hat{\mu})
 - \bar{\Psi}(x+\hat{\mu}) \gamma_\mu [P_R + u^\dagger_\mu(x) P_L] \Psi(x)\nonumber\\
+ 2\; d\; r \sum_x \bar{\Psi}(x) \Psi(x) - \frac{r}{2} \sum_{x,\mu} 
[\bar{\Psi}(x) \Psi(x+\hat{\mu})
+\bar{\Psi}(x+\hat{\mu}) \Psi(x)] + y \;\bar{\Psi}(x) \Psi(x). 
\label{cw}
\end{eqnarray}
where we have introduced a bare mass term for later use. 
When we refer to $f$-lattice quantities we use $f=1$ units and when we
refer to $b$-lattice quantities we use $b=1$ units for notational simplicity.
This should create no confusion. 

The TC path integral is then simply
\begin{eqnarray}
Z = \int_{{\cal L}_b} {\cal D}U \;\int_{{\cal L}_f} d\bar{\Psi} d\Psi \;\;\; e^{-S_{f}[\Psi,u[U]]-S_{g}[U]},
\label{pi}
\end{eqnarray}
where ${\cal D}U$ is the standard $b$-lattice gauge measure.

We still have to give a precise 
definition of the $u_\mu(x)$ in terms of the 
$b$-lattice gauge fields $U_\mu(s)$.
For a detailed explanation of how to construct such an interpolation for
the general case of a non-Abelian gauge theory in 4D, the reader is refered
to \cite{us2}. The interpolation is of course 
gauge covariant under gauge transformations on the $b$-lattice, i.e. there
exists a gauge transformation on the $f$ lattice, $\omega(x)$, such that
\begin{eqnarray}
u_{\mu}[U^\Omega](x) = \omega(x) \; u_{\mu}[U](x) \; \omega^\dagger(x+\hat{\mu}) \equiv u_{\mu}^\omega(x).
\label{gt}
\end{eqnarray}
In this way, any functional that is gauge-invariant on the $f$-lattice is 
automatically gauge-invariant under the $b$-lattice gauge transformations 
(notice that the relevant gauge symmetry is that on the $b$-lattice). 
Similarly, it is covariant under the remaining $b$-lattice symmetries ($90^\circ$ 
 rotations, translations, spatial and temporal inversions). This is  
important to ensure that Lorentz symmetry is recovered in the continuum limit
$b\rightarrow 0$. 
Finally, the interpolation has to be as smooth as possible in order
to achieve the cutoff separation. In \cite{us1}, it was shown 
that a sufficient condition to ensure that the 
gauge symmetry violations induced by the Wilson term are suppressed by $f/b$  
is that the interpolated field, in the limit $f\rightarrow 0$, describes a 
differentiable
continuum gauge field inside each $b$-lattice hypercube, and its transverse
components are continuous across hypercube boundaries. We refer to this 
property as transverse continuity. 

Now, we want to derive the equivalent Wilson-Yukawa picture. 
Following the steps of the previous section and using the property (\ref{gt}), 
the model (\ref{pi}) is exactly equivalent to,
\begin{eqnarray}
Z = \int_{{\cal L}_b} {\cal D}\Omega \; \int_{{\cal L}_b} {\cal D}U \;\int_{{\cal L}_f} d\bar{\Psi} d\Psi \;\;\; e^{-S_{f}[\Psi^\omega,u^\omega[U]]-S_{g}[U]}, 
\label{twocutwy}
\end{eqnarray}
where the $\omega$ fields are defined by (\ref{gt}) and are coupled uniquely 
through the Yukawa terms. The $\omega$ fields, which are determined by  
the $\Omega$ and $U_{\mu}$ fields and the interpolation procedure, can be interpreted as 
 smooth interpolations of the $\Omega$ fields to the $f$-lattice. 
In general the transverse
continuity property of the interpolation also implies that the
$f\rightarrow 0$ limit of $\omega$ is a continuous field. The
high Fourier modes $q > b^{-1}$ of $\omega$ are then strongly suppressed.

Now, in this picture the model has an exact gauge symmetry:
\begin{eqnarray}
U_{\mu}(x) &\rightarrow & \Phi(x) U_{\mu}(x) \Phi^\dagger(x+\hat{\mu}) \;\;\;\;\Omega(x)  \rightarrow  \Omega(x) \Phi(x)^\dagger \nonumber\\
\Psi_L &\rightarrow & \phi(x) \Psi_L(x) \;\;\;\;\;\;\;\; \Psi_R  \rightarrow   
\Psi_R,
\label{tcgt}
\end{eqnarray}
where $\phi$ is a functional of $\Phi, U_{\mu}$ defined by the 
condition,
\begin{eqnarray}
u_{\mu}[U^\Phi](x) = \phi(x) \;u_{\mu}[U](x) \;\phi^\dagger(x+\hat{\mu}).
\end{eqnarray}
 
The new gauge invariance comes at the expense of having unphysical 
degrees of freedom, $\omega$. Only if the $\omega$ fields
decouple does (\ref{tcgt}) imply the true gauge invariance of the 
original theory (\ref{pi}). 

\subsection{A Two-Dimensional Model}

In order to begin the numerical study of the models (\ref{twocutwy}), we have 
considered the simplest non-trivial model, which is a two-dimensional 
chiral $U(1)$ theory. The two-cutoff lattice action for the gauge non-invariant
picture is given by
\begin{eqnarray}
S= \sum_{s \in {\cal L}_b} {\cal L}_{gauge}+ \sum_{x \in {\cal L}_f}{\cal L}_{fermion},
\end{eqnarray}
with, 
\begin{eqnarray}
{\cal L}_{gauge} & = & -\frac{1}{g^2} {\mbox Re}\; U_{12}(s) \nonumber\\
{\cal L}_{fermion} & = & \frac{1}{2} \sum_{\mu} \bar{\Psi} \gamma_{\mu} 
[ ( D^+_{\mu} + D_{\mu}^-) P_L + (\partial^+_{\mu}+\partial^-_{\mu}) P_R] \Psi\nonumber\\
& + & y \;\bar{\Psi} \Psi - \frac{r}{2} \bar{\Psi} \sum^{2}_{\mu=1} \partial_{\mu}^+ \partial^-_{\mu} \Psi.
\label{u1}
\end{eqnarray}
where the covariant and normal derivatives are given by $D_{\mu}^+ \Psi(x) = u_{\mu}(x) \Psi(x+\hat{\mu}) - \Psi(x)$, $D_{\mu}^- \Psi(x) = \Psi(x) - u^\dagger_{\mu}(x-\hat{\mu}) \Psi(x-\hat{\mu})$, $\partial_\mu^+ = D_{\mu}^+|_{u=1}$ and 
$\partial_\mu^- = D_{\mu}^-|_{u=1}$. As explained above the $u_{\mu}$ link variables 
are interpolations of the real dynamical fields $U_\mu(s)$. The explicit
expression is given in the appendix. The field $U_{12}(s) \equiv U_1(s) U_2(s+\hat{1}) U_1^\dagger(s+\hat{2}) U_2^\dagger(s)$ is the plaquette 
variable on the $b$-lattice. 

This theory is of course anomalous. In order to 
get rid of the gauge anomaly more flavours should be introduced. However, 
we will only consider here the quenched approximation and the anomaly is not
present in this limit. The reason why already the quenched model is interesting
is two fold. Firstly, in the TC construction, it can be 
shown \cite{us1} that fermion loop corrections, which in general modify
the scalar interactions, are suppressed by powers of $f/b$. In this situation, 
it is expected that unquenching is not going to change the phase diagram in
any  essential way.
This is of course not true in the OC formulation or in an anomalous 
theory, where in general
fermion loops generate corrections to the scalar potential of $O(1)$. 
Secondly, Wilson-Yukawa models failed in giving a chiral 
theory already in this approximation, so it is important to check that 
a different picture emerges with the TC method. We will comment
on the expected effects of unquenching in an anomaly free model in section 5.

The Wilson-Yukawa picture is readily obtained from (\ref{twocutwy})
 and (\ref{u1}). The action is given by,
\begin{eqnarray}
S= \sum_{s \in {\cal L}_b} {\cal L}^{wy}_{gauge}+ \sum_{x \in {\cal L}_f}{\cal L}^{wy}_{fermion}.
\end{eqnarray}
with,
\begin{eqnarray}
{\cal L}^{wy}_{gauge} & = & -\frac{1}{g^2} Re\; U_{12}(s) \nonumber\\
{\cal L}^{wy}_{fermion} & = & \frac{1}{2} \sum_{\mu} \bar{\Psi} \gamma_{\mu} 
[ ( D^+_{\mu} + D_{\mu}^-) P_L + (\partial^+_{\mu}+\partial^-_{\mu}) P_R] \Psi\nonumber\\
& + & y \bar{\Psi} (\omega^\dagger P_R + \omega P_L) \Psi - \frac{r}{2} \left ( \bar{\Psi} \omega^\dagger P_R \sum^{2}_{\mu=1} \partial_{\mu}^+ \partial^-_{\mu} \Psi + \bar{\Psi}P_L \sum^{2}_{\mu=1} \partial_{\mu}^+ \partial^-_{\mu}  \omega \Psi \right ).
\label{u2}
\end{eqnarray}
In the case $b=f$, this model 
is identical to the one considered in \cite{smit1} for $\kappa = 0$.

As explained before, the model (\ref{u2}) is invariant under a new $U(1)_L$ 
gauge symmetry,
\begin{eqnarray}
U_{\mu}(s)& \rightarrow & \Phi(s)\; U_{\mu}(s)\; \Phi^\dagger(s+\hat{\mu})\nonumber\\
\Psi_L(x)& \rightarrow &\phi(x)\; \Psi_L(x) \nonumber\\
\omega(x)& \rightarrow &  \omega(x)\; \phi(x)^\dagger 
\label{wygt}
\end{eqnarray}
where $\phi(x)$ is obtained from $\Phi(s)$ and $U_{\mu}(s)$ from the 
condition
\begin{eqnarray}
u_{\mu}[U^\Phi](x) = \phi(x) \; u_{\mu}[U](x) \phi^\dagger(x+\hat{\mu}).
\end{eqnarray}
There is also a global $U(1)_R$ symmetry under which 
\begin{eqnarray}
\Psi_R(x) & \rightarrow &\phi_R(x) \;\Psi_R(x)\nonumber\\
\omega(x) & \rightarrow & \phi_R(x) \; \omega(x)
\end{eqnarray}
Finally, at $y=0$ there is the famous shift symmetry \cite{gp}
\begin{eqnarray}
\Psi_R \rightarrow \Psi_R + \epsilon_R,
\end{eqnarray}
which ensures that the mass of a fermion with the same quantum numbers
as the $\Psi_R$ should vanish in the limit $y\rightarrow 0$.

Since we are not interested in a spontaneously broken theory, the $\omega$ 
fields that represent the pure gauge degrees of freedom 
should decouple from the light spectrum. Thus the target 
theory we would like to recover in the continuum limit 
contains a massless charged left-handed 
 fermion and a free right-handed fermion:
\begin{eqnarray}
{\cal L}_{target} = \frac{1}{2 g^2} F_{\mu\nu} F_{\mu\nu} + \bar{\Psi} ( \not\!\!D P_L + \not\!\partial P_R ) \Psi.
\label{target}
\end{eqnarray}

As a further simplification, we are going to consider the weak
coupling limit $g \rightarrow 0$ of this model. In this limit the 
local $U(1)_L$ symmetry turns into a global one. Again this has been the 
approximation used in previous studies of standard Wilson-Yukawa models \cite{smit1}. 
In this case, since the Wilson-Yukawa model (\ref{u1})
is gauge-invariant, we can simply set $U_{\mu}=1\rightarrow u_{\mu} =1$ and what remains is a 
model of fermions coupled to the scalars $\omega(x)$. 
In this case, the $\omega$ fields depend only on the $\Omega$. 
The expression is given in the appendix, (A.9)--(A.11).
The target theory in the global limit corresponds to a theory with 
free massless fermions. It is easy to 
trace which of the fermions will couple to the gauge fields, when gauge 
interactions are switched on, by their quantum numbers under the residual 
global symmetry. 

\section{Scenarios for an effective theory}

It is instructive to review what was found in the OC formulation
of this model in ref. \cite{smit1}, in order to better understand in what
sense the two formulations differ. The model considered in \cite{smit1} 
is simply (\ref{u2}), with an additional kinetic term for the scalar 
fields:
\begin{eqnarray}
{\cal L}_{scalar} = - \kappa \sum^{2}_{\mu=1} [\;  \Omega^\dagger(s)\; \Omega(s+\hat{\mu}) + \Omega^\dagger(s+\hat{\mu})\; \Omega(s) \;].
\label{sca}
\end{eqnarray}
The scalar theory (\ref{sca}), for $f=b$ (OC), is the well-known
XY model, which is known to have a phase transition 
at $\kappa=\kappa_c \approx 
0.56$, which separates a vortex phase (VX) for $\kappa < \kappa_c$ 
with finite scalar correlation length from a spin-wave (SW) phase, 
where the scalar correlation length is infinite \cite{XY}. In the quenched 
approximation, the fermions do not modify the scalar interactions, so 
this scalar dynamics is independent of the scalar-fermion couplings $r$ and $y$.
Since we
are interested in a theory with no spontaneous symmetry breaking, we 
should consider the VX phase ($\kappa < \kappa_c$). Without any tuning 
towards the critical line, the scalars have masses of the order of the 
cutoff. 

The scalar correlation length decreases as $\kappa$ is lowered, in agreement 
with the FNN conjecture, since 
the $\Omega$ fields interactions are controlled by $\kappa$. Alternatively, 
interpreting 
$\Omega$ fields as the pure gauge transformations in a $U(1)$ gauge model, 
$\kappa$ measures the strength of the breaking of gauge invariance at the 
cutoff 
scale. According to the FNN argument, at distances much larger than the scalar 
correlation length, the theory should have an effective gauge symmetry. 

In this model, there are also interactions between the scalars
and the fermions through the Wilson term in (\ref{u2}). This Wilson term 
is not small for large $r, y$ and it induces 
large couplings between the fermions and the scalar fields. However,
 at least naively it seems that the FNN argument can still be applied
since, in the quenched approximation, the scalar
correlation length is controlled 
only by $\kappa$ and, for $\kappa \ll \kappa_c$, it is of the order of the cutoff. 
At distances that are large compared to 
this correlation length, the physics can be described by an effective
theory where the scalar fields are integrated out. The effective Lagrangian
should be local, because it comes from the integration of very heavy particles 
and it is exactly gauge-invariant, due to the exact symmetry (\ref{wygt}). 
Of course, this is not very useful, because it does not tell us what the
light fermion spectrum is or if there is any. 

When the Yukawa couplings 
$r$ and $y$ are small, perturbation theory in these couplings 
 can be applied and it is clear that the asymptotic fermion states are the 
original ones in the Lagrangian. Since they have chiral quantum numbers, they
should be massless in the FNN regime, already at finite lattice spacing, to 
ensure gauge invariance. 
This is precisely what OC studies found in the VX (or PM in 4D)
weak Yukawa phase, VXW (PMW). In the weak coupling regime, fermion masses are proportional to the 
scalar vev, i.e.,  
\begin{eqnarray}
m \sim \langle \Omega \rangle,
\label{pth}
\end{eqnarray}
thus vanishing in the symmetric VX phase ($\kappa < \kappa_c$).
However, as is well-known, the spectrum is doubled in this phase, because
both doubler and light mode masses behave as (\ref{pth}).

For large Yukawa couplings, the problem is much more difficult. 
Strong coupling expansions and numerical simulations in OC constructions 
have shown \cite{wy, smit1} that in fact 
there is a free massive fermionic spectrum, which gets light when 
$y$ is taken to zero (the doubler masses however 
do not go to zero in this limit). 
This is the strong VX (PM) Yukawa phase, VXS (PMS).
In this phase, the fermion masses are proportional to a globally symmetric
condensate of the form,
\begin{eqnarray}
m \sim \langle {\mbox Re} [ \;\Omega(s) \;\Omega^\dagger(s+\hat{\mu}) ]  \rangle^{-1/2}
\label{stro}
\end{eqnarray}
which is non-zero in the VX phase. 
Although the doublers can easily be decoupled in this phase, the light 
massive fermions were found to transform vectorially under the 
global group. Obviously they have to be different from the ones appearing
in the Lagrangian (\ref{u2}), since these have chiral quantum numbers. 
Notice that according to FNN, this is in fact the only possibility: if the 
light fermions are massive at finite lattice spacing, they must be 
vector-like in order to satisfy gauge invariance. 
The transition line between the VXS and VXW phases was found to be, for this model at $y+ 2\; r \simeq 1$, independently of the individual values of $y$ and $r$ \cite{smit1}. 

 The existence of a strong Yukawa phase implies that, even though the 
global chiral symmetry is not broken (in 2D this has to be the case 
even in the SW phase according to Mermin-Wagner-Coleman theorem), mass 
terms appear. 
The solution of this apparent contradiction is that the fundamental massive
fermions are not the ones in the original Lagrangian (\ref{sca}), but 
new fermionic composites made out of the original ones and
the scalar fields, which are expected to form in the strong 
coupling regime \cite{smit1,witten}. It is easy to see 
that the Dirac fermions,
\begin{eqnarray}
\Psi^{(n)} \equiv  \omega \; \Psi_L + \Psi_R \nonumber\\
\Psi^{(c)} \equiv  \Psi_L + \omega^\dagger \Psi_R, 
\label{composites}
\end{eqnarray}
transform vectorially under the global group. Notice that 
for $f/b = 1$, $\Omega = \omega$. $\Psi^{(c)}$ is charged
under the $U(1)_L$ and $\Psi^{(n)}$ is charged under the $U(1)_R$. Since
it is conventional to gauge the $U(1)_L$, these 
composites are often called charged and neutral, respectively. It is clear
that mass terms for these fields do not break the global symmetry. We will 
call physical the Dirac field made out of the original fields in 
(\ref{sca}), i.e. $\Psi^{(p)} \equiv \Psi_L + \Psi_R$, since it is 
the one we expect to find in the light spectrum according to (\ref{target}). 

When the couplings $y$ and $r$ are large ($O(1)$), the fermion-scalar couplings
in (\ref{u2}) are strong enough to favour the formation of the  mirror 
composites, $\omega \; \Psi_L$ and $\omega^\dagger \Psi_R$. 
The numerical simulations of \cite{wy,smit1} showed that the
strong phase could be reached by making $r$ large while 
keeping $y$ arbitrarily small, as long as the condition $y + 2\; r > O(1)$ was
satisfied. 
In fact, the numerical results for the neutral propagator ($\Psi^{(n)}$) 
were nicely reproduced by the first-order 
corrections in the hopping parameter expansion ($\frac{1}{y+ 2 r}$) \cite{wy}, 
which does not distinguish between the regions $r \approx y = O(1)$ and 
$y \ll r = O(1)$. To first order in this expansion, the neutral field is 
a free Wilson fermion. The 
Lagrangian for the neutral field is given by,
\begin{eqnarray}
{\cal L}^{(n)}  & = & \frac{1}{2} \sum_{\mu} \bar{\Psi}^{(n)} \gamma_{\mu} 
[\; \omega\; ( D^+_{\mu} + D_{\mu}^-)  \omega^\dagger P_L+ (\partial^+_{\mu}+\partial^-_{\mu}) P_R \;] \Psi^{(n)}\nonumber\\
& + & y \;\bar{\Psi}^{(n)} \Psi^{(n)} - \frac{r}{2}  \bar{\Psi}^{(n)} \sum^{2}_{\mu=1} \partial_{\mu}^+ \partial^-_{\mu}  \Psi^{(n)}  
%\equiv \bar{\Psi}^{(n)} \;M^{(n)} \Psi^{(n)}.
\label{neutral}
\end{eqnarray}
The zeroth-order propagator is given by, 
\begin{eqnarray}
S^{(0)}(q) = \frac{ - i \sum_\mu \gamma_\mu \;\sin q_\mu P_R + M(q)}{M(q)^2}
\end{eqnarray}
with $M(q) \equiv 1 - \alpha\; r \sum_\mu \cos q_\mu$ and $\alpha \equiv 1/(y + 2  r)$. At higher orders, the
propagator can be writen in general as
\begin{eqnarray}
S^{(n)}(q) = \frac{1}{ i \sum_\mu \gamma_\mu \sin q_\mu (f^{(n)}_R(q) P_R + f^{(n)}_L(q) P_L)+ M^{(n)}(q) + M^{(n)}_5(q) \gamma_5}.
\label{prop}
\end{eqnarray}
At first order in $\alpha^2$, one gets \cite{wy}
\begin{eqnarray}
f^{(1)}_R(q) & = &  1 \;\;\;\;\;f^{(1)}_L(q)  =  (\alpha z)^2 \nonumber\\
M^{(1)}(q) & = &  M(q)  \;\;\;\;\;M_5^{(1)}(q) = 0
\label{hop1} 
\end{eqnarray}
where $z^2$ is the well-known condensate
\begin{eqnarray}
z^2 \equiv \;\langle {\mbox Re}( \omega(x) \omega^\dagger(x+\hat{\mu}) ) \rangle_{\Omega}.
\end{eqnarray}
In the strong Yukawa phase, $z^2$ plays the same role as the vev in 
the broken phase. However, 
it is  invariant under the global chiral symmetry, as opposed to the normal vev. This is of course the reason why it induces fermion masses in the 
symmetric phase. It is easy to see that the propagator defined by
 (\ref{prop}) and (\ref{hop1}) is 
equivalent to a free Wilson propagator with, 
\begin{eqnarray}
y_{eff} = \frac{y}{z}, \;\;\;\;\;\; r_{eff} = \frac{r}{z},\nonumber
\end{eqnarray}
\begin{eqnarray}
Z_{LR} = z^{-2},\;\;\;Z_{RL} = 1,\;\;\;Z_{LL} = Z_{RR} = z^{-1}.
\label{effcou}
\end{eqnarray}

The Lagrangian for the charged fermion 
is given by, 
\begin{eqnarray}
{\cal L}^{(c)}  & = & \frac{1}{2} \sum_{\mu} \bar{\Psi}^{(c)} \gamma_{\mu} 
[  ( D^+_{\mu} + D_{\mu}^-)  P_L+ \omega^\dagger ( \; \partial^+_{\mu}+\partial^-_{\mu}) \;\omega\;P_R\; ] \Psi^{(c)}\nonumber\\
& + & y \;\bar{\Psi}^{(c)} \Psi^{(c)} - \frac{r}{2}  \bar{\Psi}^{(c)} \omega^\dagger \sum^{2}_{\mu=1} \partial_{\mu}^+ \partial^-_{\mu}  \omega \Psi^{(c)}  
\label{charge}
\end{eqnarray}
In this case, the numerical results 
were not well reproduced by the hopping expansion. On the contrary, there was 
strong evidence that this fermion is not a free asymptotic 
state \cite{gpr}, but most probably a two-particle state, $\omega^\dagger 
\Psi^{(n)}$.

In summary, OC studies found that both the strong and weak Yukawa phases
failed in giving a true chiral spectrum. As we have seen, these results are
 in nice agreement with the FNN mechanism. The scalar fields decouple at 
large distances and an effective gauge-invariant theory remains. However 
the gauge symmetry can be realized either through a massive vector-like fermion 
spectrum or through a doubled massless one. FNN does not exclude the 
possibility
that composite fermions exist at large distances nor that there is doubling.
  We will now argue that 
in the two-cutoff formulation (\ref{u2}), there is a richer structure 
in the strong Yukawa phase, depending on the individual values of $y$ and $r$.

Clearly the only difference between OC and TC is in the scalar dynamics. 
In the TC case we have set $\kappa = 0$ for simplicity, so we are deep 
in the VX phase. In OC
studies the fermion spectrum has not been analysed in detail in 
this extreme region of phase space. Probably one of the reasons is that it
is much more demanding numerically. The other reason is that in the unquenched theory  fermion loops
introduce large (typically $O(1)$) corrections to $\kappa$, so that in the full 
theory it makes no sense to consider the case $\kappa=0$ unless some fine-tuning is done. In the TC case, as we will see, fermion loops introduce 
only a small correction to $\kappa$, provided there are no gauge anomalies. 

The main difference between the two formulations is the fact that, 
due to the interpolation, the 
high Fourier modes of the scalar field are suppressed. 
Let us consider a region of phase space with $y$ arbitrarily small and 
$r =O(1)$. We use $f=1$ units unless stated otherwise. 
According to the OC studies this would correspond to the 
strong phase, where the spectrum is vector-like. However, for these 
couplings, the 
light lattice mode (i.e. $p \rightarrow 0$) is coupled to the scalar field 
only through a perturbatively small 
$y$ Yukawa coupling, and a higher-dimensional operator. At small
lattice momentum, the effect of this operator is small. Notice that this is 
true only for TC. In the OC case, the Wilson term also induces a large $O(r)$ 
coupling between the light mode and the doublers, so it is not possible to treat
it in perturbation theory. In the TC case, this coupling is small, because
it involves high-momentum modes of the scalar field ($\sim f^{-1}$), which are suppressed. 
On the other hand, for the doubler poles,
$p \rightarrow \pi$,  the Wilson term 
is effectively a Yukawa coupling of strength $O(r)$, in both OC and TC 
formulations. 
In this situation, we expect that 
at small
enough $y$, the light mode enters into a weak phase, because the strength
of the Yukawa couplings is small at $p \rightarrow 0$, and there is no 
reason to believe that the formation of the mirror state is dynamically
favoured. On the other hand, 
the doubler modes could remain in the strong phase, since for them the Yukawa
couplings are of $O(1)$. This of course would be a most
desirable scenario, since the doublers would be decoupled from the light 
spectrum, while the 
light modes would remain massless without any extra tuning, in the same way as
they are in the weak Yukawa phase. In other words, the light fermion spectrum 
would be 
chiral and not doubled. For this reason we will refer to this phase as 
the chiral phase. 
Clearly, the fact that the scalar field momenta cannot get as large
as the fermion cutoff (the doubler momentum) is essential in this argument.

If a new phase of the kind discussed above exists at 
small $y$, it cannot be described 
by the lowest orders of the hopping parameter expansion and it is reasonable
to 
expect that this expansion should signal its failure to converge at small $y$. 
The first-order hopping result (\ref{hop1}) corresponds to the resummation of a set of one-loop tadpole
diagrams, which is a geometric series in the quantity $(\alpha z)^2 \sum_\mu (\sin q_\mu)^2/M(q)^2$. If the corrections that are neglected at the next order are of the same order of magnitude as the second term in this series, the 
hopping expansion is expected to be a good approximation only 
if  $(\alpha z)^2 \sum_\mu (\sin q_\mu)^2/M(q)^2 \ll 1$, which is a momentum-dependent condition.   
At small momentum, $M(q) \rightarrow 1 - 2 \alpha r \sim y / 2 r$ and 
the convergence condition would require $z/(y L) \ll 1$, where $L$ is the 
lattice size. For small enough $y$ the condition is clearly not satisfied, since $z^2$ is independent of $y$ and, as we will see, in the TC case, it is $\sim 1$. 

In the region of small $y$ and at small momenta, we expect to find a 
chiral phase, in which the light fundamental fields are those in the original
Lagrangian (\ref{u2}) that we have called physical.
This phase is clearly more difficult to analyse than both the strong and 
weak phases, since doublers and light
modes behave in a very different way, even though they are excited by 
the same field. At large momenta, the hopping expansion is expected
to be a good approximation, while at small momenta weak perturbation theory
should be a good approximation. 
The exact result for the propagator of the physical 
fields in momentum space
is then expected to be an interpolation between the hopping result (\ref{hop1}) 
at large momenta and the weak expansion result at small momenta. 

At small momentum and to zeroth order in $y$ and $r$:
\begin{eqnarray}
(S^{(p)}(q))^{0} = (i \sum_\mu \;\gamma_\mu \sin q_\mu)^{-1}, \;\;\;\;   q\rightarrow 0.
\end{eqnarray}
Higher-order corrections can induce a renormalization of the fields so 
in general we expect that the propagators will be well reproduced by 
\begin{eqnarray}
S^{(p)}(q)_{LR} = S^{(c)}_{LR} = \frac{Z_L}{i \sum_\mu \gamma_\mu \sin q_\mu}, \;\;\;\;\;
S^{(p)}(q)_{RL} = S^{(n)}_{RL} = \frac{Z_R}{i \sum_\mu \gamma_\mu \sin q_\mu} \;\;\; q_\mu \rightarrow 0.
\label{chiralph}
\end{eqnarray}

The $LL, RR$ components of $S^{(p)}$ are zero, because there
is no chiral symmetry breaking. However all the remaining components of the
neutral and charged propagators, i.e. $S^{(n)}_{RR,LL,LR}$ and $S^{(c)}_{RR,LL,RL}$ are non-zero. In weak perturbation theory, they correspond to 
multiparticle states made of various combinations 
of the massless fields, $\Psi^{(p)}_R$,  
$\Psi^{(p)}_L$ and the scalar $\omega$. For instance, at large times, 
we expect that 
the propagation of $\Psi^{(n)}_L$ and $\Psi^{(c)}_R$ is dominated by two-particle states, i.e.
\begin{eqnarray}
S^{(n)}_{LR}(x,y) \sim S^{(p)}_{LR}(x,y)\; G_{\omega}(x,y),\;\;\; 
S^{(c)}_{RL}(x,y) \sim S^{(p)}_{RL}(x,y)\; G_{\omega}(x,y),
\label{chiexp}
\end{eqnarray}
where $G_{\omega}$ is the scalar propagator.
Since these propagators involve  scalars, which have masses of the order
of the boson cutoff, they should decouple from the light 
spectrum. 

At large momentum, a better approximation is provided by the 
strong coupling expansion. In this case the physical propagators 
$S^{(p)}_{RL}$ and $S^{(p)}_{LR}$ combine with the remaining neutral 
and charged components to form massive Dirac fields. Thus we do not expect
light doubler poles. The physical propagators at large momentum should behave as
\begin{eqnarray}
S^{(p)}(q)_{RL,LR} =  Z_{R,L} \frac{- i \sum_\mu \gamma_\mu \sin q_\mu  + M_{R,L} }{\sum_\mu (\sin q_\mu )^2 + M_{R,L}^2},\;\;\;\;\;\;  q_\mu  \rightarrow \pi.
\label{chima}
\end{eqnarray}
 
To summarize, the expected phase diagram in the symmetric phase, 
in the TC formulation, has three regions (not necessarily separated by 
phase boundaries). At small $y, r \ll 1$, there is a weak 
phase where all fermions are massless, doublers and light lattice modes. 
The effective theory in this regime is different from the target theory
(\ref{target}) in that there is a doubling: for each mode
with a $P_L$ coupling to the gauge field, there is one with a $P_R$
coupling. At large $y, r = O(1)$, there is a strong phase where all fermions
are massive, although chiral symmetry is not broken. The real asymptotic
fermion state is a singlet fermion (the charged one is possibly just a 
two-particle state, as favoured by the OC studies \cite{gpr}), which only 
couples to the gauge fields through higher-dimensional operators. This is
also far from (\ref{target}). Even if the charged fermion existed as 
a free fermion in this phase, it would couple vectorially to the gauge 
fields, so again the continuum effective theory would be different from 
the target theory. Fortunately, there is a third region at $y \ll r = O(1)$, 
which was not present in OC models, 
where the doublers are decoupled from the light mode, but the latter 
is only weakly coupled to the scalar field and remains in the weak phase. 
This chiral phase is expected to give rise to the target theory defined
by (\ref{target}). Our main purpose in the following is to show clear
evidence of the existence of this phase in the TC formulation. 

\section{Numerical results}

We have computed the fermion propagator for the different 
fields $\Psi^{(p)}$, $\Psi^{(n)}$ and $\Psi^{(c)}$, by inverting the appropriate
fermion matrices on configurations of the $\omega$ fields that 
were obtained by interpolating a 
set of completely random $\Omega(s)$ configurations, in the way described in the appendix.
This corresponds of course to the case $\kappa = 0$. For the fermion 
fields, we have used periodic boundary conditions in the spatial direction and 
antiperiodic in the temporal direction. This is necessary to study a regime 
in which we expect to find massless fermions.  On the other hand, the scalar
fields $\Omega$ are periodic in both directions. 

With two lattices, 
there is some freedom in the choice of the ratios $b/f$ and $L/b$ and 
it is not obvious how the continuum limit should be defined in this case. 
Clearly, the two cutoffs must be sent to infinity, i.e. $f, b\rightarrow 0$.
When gauge interactions are switched on, we expect to have 
a mass gap ($\xi$), so that the continuum limit is defined by, 
\begin{eqnarray}
\frac{b}{\xi} \rightarrow 0
\end{eqnarray} 
This however might require a simultaneous tunning of $f/b \rightarrow 0$ to
ensure that we do not exit the FNN regime.
In \cite{us1}, it was argued that a conservative way of taking 
the continuum limit without exiting
 the FNN regime (i.e. ensuring that fermion loops contributions to the 
scalar dynamics are suppressed) would be 
\begin{eqnarray}
\frac{f}{b}\cdot \frac{L}{b} \rightarrow 0,\;\;\;\;\;\; \frac{b}{\xi} \rightarrow 0;
\end{eqnarray}
however, it is not possible from theoretical considerations to 
determine quantitatively how these two limits should be correlated. 
To understand this issue a full simulation with gauge interactions 
would be required since, in the 
simplified model we are considering here, there is no physical mass gap. 

We have used
units of $f=1$ throughout the paper, unless stated otherwise, for notational convenience. 
However, it is important to remember that the boson cutoff is $1/b$ and thus 
in $f$-lattice units, masses of $O(f/b)$ are of the order of the gauge boson
and scalar cutoffs.
Finite-volume effects can be studied by 
keeping the couplings fixed in $b$ units, while increasing the ratio
$L/b$. This might require a simultaneous change in the ratio
$b/f$ to ensure that we remain in the FNN regime. We will refer to 
the different lattice sizes with the notation $(L/b)_{b/f}$. For 
convenience we have 
used square lattices, so that $L/f$ is the same in both space and time directions. 

\subsection{Scalar Dynamics}

The main difference with OC simulations is clearly in the scalar 
dynamics. The scalar propagator
\begin{eqnarray}
G_{\omega}(p) = \langle \;\;\frac{1}{L^2} \sum_{x,y} \omega^\dagger(x) \;\omega(y) e^{i p (x- y)}\;\;\rangle
\end{eqnarray}
 is quite different from the one corresponding
to a free massive boson field. In Fig. 1 we show the scalar propagator in 
momentum space
for an  $8_8$ lattice. The two sets correspond to $G_\omega(q_x, 0)$ and
$G_\omega(0, q_y)$ respectively. They are compatible within statistical 
errors, as expected from the lattice $90^\circ$ rotational symmetry. 
At small momentum the propagator can be fitted
to the free ansatz (dashed line in Fig. 1):
\begin{eqnarray}
G_{\omega}(q) = \frac{Z_\omega}{\hat{q}^2 + m^2_{\omega}},
\label{fa}
\end{eqnarray}
with $\hat{q}^2 \equiv 2\; (1 - \cos q)$. In this case, the fitted scalar 
mass is roughly $m_\omega = 1.7$ in $b$ units. This is in agreement
with the naive expectation that this mass is related to the boson cutoff 
 $1/b$.  
\begin{figure}
\begin{center}
\mbox{\epsfig{file=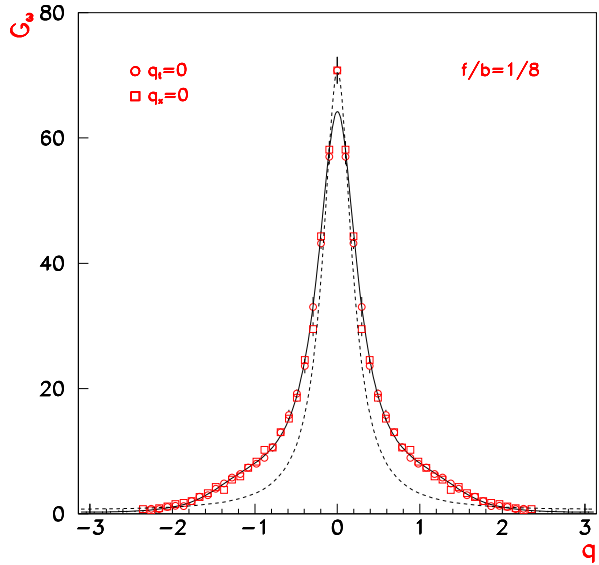,width=3.5in,height=3.5in}}
\end{center}
\caption[]{Scalar Propagator in momentum space in a $8_8$ lattice. The
dash line is the fit to the free ansatz (\ref{fa}) at small momenta.
The solid line corresponds to a fit to the inverse of a
 fourth order polynomial in $\hat{q}^2$.}
\label{scalar}
\end{figure}
In fact, if the interpolation was strictly local (i.e. the interpolated
field in each $b$ plaquette depends only on the surrounding links), the correlation length 
of the $\omega$ fields at distances larger than $2 b$ would vanish. In this case, it would not be appropiate to talk about a scalar mass. However, due to 
the fact that the group is compact, a smooth interpolation cannot 
be strictly local \cite{us2}. Empirically we find that the non-locality of the 
interpolation is nevertheless consistent with an exponential decay of the scalar propagator 
at large distances ($> 2 b$), with a mass of the order of the boson cutoff. 

\begin{figure}
\begin{center}
\mbox{\epsfig{file=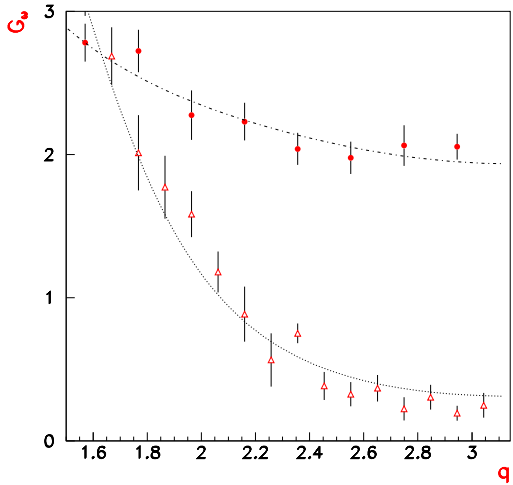,width=3in,height=3in}}
\end{center}
\caption[]{Scalar propagator in the large-momentum region for two values
of $f/b=1/8$ (triangles), $1/4$ (circles). The curves are fits to  the inverse 
of a fourth-order polynomial in ${\hat{q}}^2$ in the whole momentum range.}
\label{sca2}
\end{figure}

At large momentum, however, the $\omega$ propagator differs considerably 
from the free ansatz. Since the  interpolation is smooth, the momentum modes higher than $q \sim f/b$ (boson cutoff) are expected to be 
suppressed for small $f/b$. Empirically, the 
inverse propagator at large momenta can be fitted for these lattice sizes to a
polynomial in $\hat{q}^2$ of fourth order (solid line in Fig. 1). In Fig. \ref{sca2}, 
we show the large-momentum behaviour of the scalar propagator for two 
different values of the two-cutoff ratio $f/b$. Clearly, the 
high-momentum suppression increases as this ratio is decreased, as expected.

Concerning the condensates, there is also some difference with the OC case. 
The chirally breaking condensate,
\begin{eqnarray}
v = \langle\; \frac{1}{L^2} \sum_x \omega(x)\; \rangle
\end{eqnarray} 
is expected to be zero, since we are in the  symmetric phase. Numerically, 
it is zero within statistical errors. 

The condensate relevant for the strong phase,
\begin{eqnarray}
z^2 \equiv \langle \frac{1}{2 L^2} \sum_{x,\hat{\mu}} {\mbox Re}[\;\omega(x) \omega^\dagger(x+\hat{\mu})\;] \rangle, 
\end{eqnarray}
is plotted in Fig. 3 for two  lattice sizes of $L/b =8, \;4$, as a function of the ratio
$f/b$. It is easy to show from the explicit expression of the interpolated 
$\omega$ fields that,
\begin{eqnarray}
\lim_{f/b\rightarrow 0} \;\; z^2 = \;\; 1
\end{eqnarray}
for fixed $L/b$. 
This is in contrast to what happens in the  OC formulation, where 
$z^2 = 0$ for 
$\kappa = 0$. This important difference between the OC and TC cases is not
strange, since $z^2$  gets important contributions from short 
distances (as opposed to $v$, which only depends on the long-distance 
scalar dynamics) and the short-distance scalar dynamics is drastically
 modified by the interpolation. 
\begin{figure}
\begin{center}
\mbox{\epsfig{file=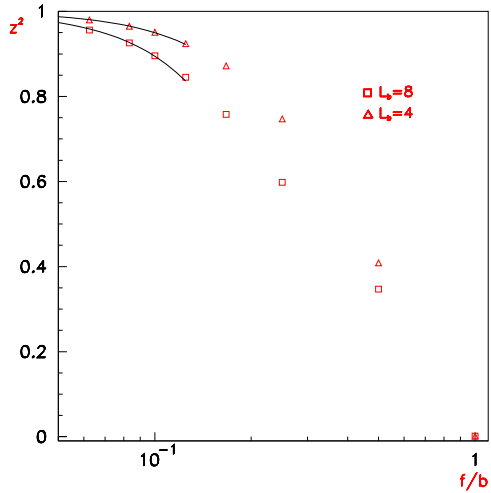,width=3in,height=3in}}
\end{center}
\caption[]{$z^2$ condensate for two lattice sizes $L/b=8, 4$ as a function of 
the ratio $f/b$. The lines are quadratic fits.}
\end{figure}

One important question is what the volume dependence of these results is. 
Since the $\Omega$ fields are random at the scale $b$ we do not expect 
an important dependence on $L/b$ \footnote{Unfortunately this
 cannot be made into a 
rigorous statement because of the winding fields $N(s)$, which can grow with
$L/b$. See the appendix.}. 
We have computed the scalar propagators for various lattice
sizes, and we find that the scalar ``mass'' (as obtained from the momentum fit
(\ref{fa})) is proportional to the ratio $f/b$, as
expected. This means that the scalar mass in $b$ units does not depend
on $f/b$. On the other hand, there is no clear dependence on $L/b$. This is 
shown in Table 1. In all the cases the scalar masses are of $O(1)$ in $b$-lattice units. 
\begin{table}
\begin{center}
\begin{tabular}{|c|c|c|c|c|c|c|}\hline
$f/b$ & $L/b$ & $m_{\omega}$ ($b$ units)  \\ \hline  
1/8 & 6 & 1.70(6) \\ 
 & 8 & 1.72(6)  \\
 & 10  & 1.75(15) \\
\cline{1-3} 
1/4 & 8 & 1.69(6) \\
   & 10 & 1.54(10) \\
\hline
\end{tabular}
\caption[]{Scalar mass in units of $b$ for various lattice sizes from momentum
fits (\ref{fa}).}
\end{center}
\end{table}

The volume dependence of the condensate $z^2$  is shown in Fig 3. 
For fixed $f/b$, 
there is some dependence on the volume\footnote{This effect is also due to 
the winding fields $N(s)$.}. In order to keep $z^2$ 
constant, the infinite volume limit must be taken in a correlated way with the 
ratio $f/b$. A sensible way of taking
 the infinite volume limit might be by following lines of constant $z^2$ 
in the $(f/b, b/L)$ plane. From Fig. 3 we see that, according to this rule, $f/b$ does not need to decay as fast as $b/L$.

\subsection{Fermion Spectrum}

The fermion propagators have been obtained with the conjugate-gradient 
method for several lattice sizes and 
$y$ couplings. Since we are only interested in the regime where doublers
are decoupled, we have kept the Wilson coupling fixed at $r=1$. According 
to the OC phase boundary $y + 2 r = O(1)$, we are then in the strong phase. 
As we will see, the strong phase in the TC has more structure, depending on 
the value of $y$. 

The various propagators are given by
\begin{eqnarray}
S_{ij}^{a}(t) =  \langle \sum_{x_1,y_1} \;\; \Psi_i^{(a)}(x) \bar{\Psi}_j^{(a)}(y) \;\; e^{i p_1 (y_1 - x_1)} \;\; \rangle_{\Omega}  , \;\;\;\;\; t = |y_2 - x_2|,
\end{eqnarray}
where $t = 1,...,L$, the index $a$ refers to the neutral (n), charged (c) or 
physical (p) fermion for $p_1=0$, and the corresponding spatial doublers (nd), (cd) and (pd) for
$p_1=\pi$. $i, j = R, L$ for the different chiralities. 
For every inversion the time slice at the 
origin $x_2$ is chosen randomly. The number of sampled scalar 
configurations is typically of $O(2-5 \ 10^2)$. 

As in the OC construction, invariance of the action 
(\ref{neutral}) under $PT$, i.e. a spatial and temporal inversion\footnote{Notice that $T$ is not equivalent to time reversal in the continuum.}, implies
\begin{eqnarray}
S^{a}(t) = - \gamma_5\; S^{a}(T-t)\; \gamma_5,
\label{tsym}
\end{eqnarray}
which means that there is antisymmetry under $t \rightarrow T-t$ for 
the chirality-breaking components, RR and LL, and symmetry for the chirality-preserving ones,
RL and LR. Notice that there is no symmetry under $T$ or $P$, since the 
action explicitly breaks parity. 

In order to check that these symmetries are well satisfied we have 
monitored the behaviour $t < T/2$ and $t > T/2$ of the propagators. The 
expected $PT$ symmetry is well satisfied within statistical errors.  
If the low lying spectrum consists of free massive
fermions, $P$ and $T$ symmetries are also expected to be satisfied
effectively, while if the spectrum is chiral $P$ and $T$ are broken explicitely.  

\subsubsection{Large-$y$ Regime}

At large $y \gg f/b$, we find that all the fermions are massive, as 
was also found in OC studies. Of course this phase has no physical interest
since the fermionic states are more massive, $m \sim y$, than the scalars and
 this implies that there is no FNN regime to look at. All the particles, 
scalars and fermions, have masses of the order of the cutoff. 
In any case, it is interesting to look at this regime in order to compare
with OC studies and to confirm that also in the two-cutoff formulation, for
large Yukawa couplings, fermions are massive even if chiral symmetry is not 
broken.
\begin{figure}
\begin{center}
\mbox{\epsfig{file=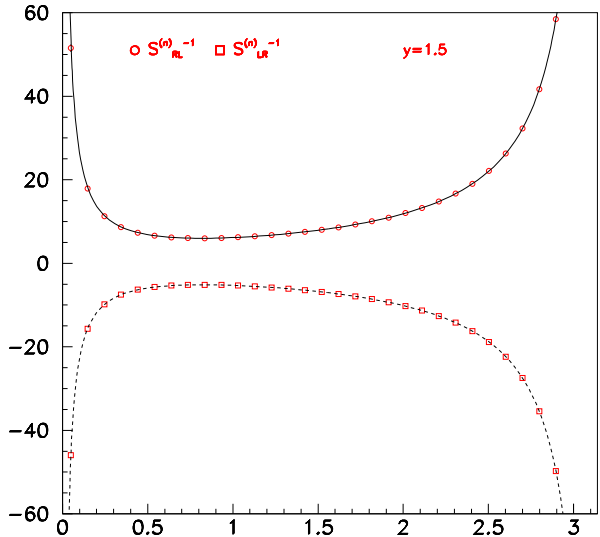,width=3.5in,height=3.5in}}
\end{center}
\caption{$(S^{(n)}_{RL,LR})^{-1}$ at $y=1.5$ and a lattice size of $8_8$.
 The solid lines correspond to the fits to the free Wilson fermion formulae
of eq. (\protect\ref{fit}). The parameters of the fits are given in Table 2.}
\label{fig3}
\end{figure}
Figure \ref{fig3}  shows the inverse neutral propagators in momentum space 
for $y=1.5$ in an $8_8$ lattice. The solid lines correspond to the 
fits to the free Wilson formula (for zero spatial momentum), i.e.
\begin{eqnarray}
{S^{(n)}_{LR,RL}(q)}^{-1} = \left [\;Z_{L,R} \;\frac{\sin q }{(\sin q)^2 + ( y_{eff} + r_{eff} (1- \cos q))^2} \; \right ]^{-1}\nonumber\\
{S^{(n)}_{LL,RR}(q)}^{-1} = \left [\;Z_{LL,RR} \;\frac{ y_{eff} + r_{eff} ( 1- \cos q  )}{(\sin q)^2 + ( y_{eff} + r_{eff} (1- \cos q ))^2} \; \right ]^{-1}
\label{fit}
\end{eqnarray}
with $Z_{LL,RR} = (Z_{L} Z_{R})^{1/2}$.
 All propagators can be fitted with a very good $\chi^2$
to the free Wilson propagator. The hopping expansion is a very
good approximation at large momentum ($q > \pi/b$) 
as can be seen in Table 2. We present
the results for $y_{eff}$ and $r_{eff}$ of the fits to (\ref{fit}), excluding the points
near zero momentum, together with the result of the hopping expansion to 
first order (\ref{effcou}). If the points near zero momentum are included
in the fit the $\chi^2$ increases considerably, indicating that the free propagator (\ref{fit}) is
not a good approximation in the low-momentum region. 

\begin{figure}
\begin{center}
\mbox{\epsfig{file=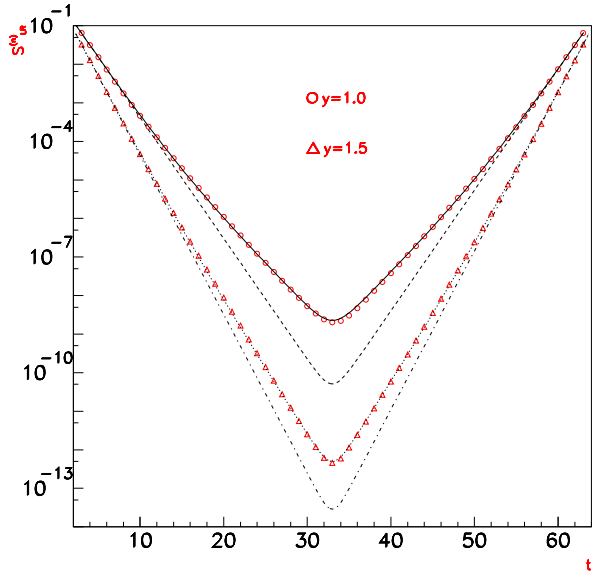,width=3.5in,height=3.5in}}
\end{center}
\caption[]{$S^{(n)}_{LR}$ propagators in time for $y = 1, 1.5$ and a $8_8$ lattice. The lines
following the data points are the result of a two exponential fit.
The dotted and solid lines, which deviate at large times, are the results of the momentum fits of Table 2.}
\label{fig4}
\end{figure}

This effect is more clearly seen in 
the behaviour of the propagators at large times. 
\begin{table}
\begin{center}
\begin{tabular}{|c|c|c|c|c|c|c|c|}\hline
y & Prop & & RR & RL & LR & LL & Hopp.\\ \hline
1. & $S^{(n)}$ & Z & 1.084(5)& 1.006(23) & 1.175(19)& 1.086(5) & \\  
   &           &$y_{eff}$ &1.090(6)& 1.081(18) & 1.087(11) & 1.093(6) & 1.085\\
   &           &$r_{eff}$ &1.081(6)& 1.090(11) & 1.083(8) & 1.082(5) & 1.085\\ 
\cline{2-8}
   & $S^{(nd)}$ & Z &1.084(5)& 0.992(26) & 1.176(10)& 1.085(5) & \\  
    &           & $y_{eff}$ &1.085(5)& 1.064(14) & 1.085(5) & 1.086(5) & 1.085\\
    &           & $r_{eff}$ &1.084(5)& 1.086(14) & 1.084(5) & 1.085(5) & 1.085\\
\hline
1.2 & $S^{(n)}$ & Z &1.082(5)& 0.997(23) & 1.172(11)& 1.085(5)& \\  
    &           & $y_{eff}$ &1.304(5)& 1.289(22) & 1.301(6) & 1.307(6)& 1.302\\
    &           & $r_{eff}$ &1.080(5)& 1.085(14) & 1.082(5) & 1.082(5) & 1.085\\
\cline{2-8}
   & $S^{(nd)}$ & Z &1.084(5)& 0.969(32) & 1.175(10)& 1.085(5) & \\  
    &           & $y_{eff}$ &1.301(6)& 1.266(20) & 1.301(6) & 1.302(6) & 1.302\\
    &           & $r_{eff}$&1.084(5)& 1.072(20) & 1.084(5) & 1.085(5) & 1.085\\ 
\hline
1.5 & $S^{(n)}$ & Z &1.084(4)& 1.009(15) & 1.175(8)& 1.085(3) & \\  
    &           & $y_{eff}$&1.630(6)& 1.624(13) & 1.628(5) & 1.631(5)& 1.627\\
    &           & $r_{eff}$ &1.082(4)& 1.093(7) & 1.083(4) & 1.083(3) & 1.085\\ 
\cline{2-8}
   & $S^{(nd)}$ & Z &1.085(3)& 0.991(25) & 1.178(13)& 1.085(3) & \\  
    &           & $y_{eff}$ &1.628(5)& 1.608(19) & 1.629(10) & 1.628(5)& 1.627\\
    &           & $r_{eff}$ &1.085(3)& 1.083(13) & 1.085(6) & 1.085(3) & 1.085\\ 
\hline
\end{tabular}
\caption[]{Fits of the neutral propagators (light and doubler) in momentum space to the free Wilson formula for large $y$ in an $8_8$ lattice ($z^{-1} = 1.085$).}
\end{center}
\end{table}
In Fig. \ref{fig4}, we present the $LR$ component of the neutral 
propagator for two large values of $y$ together with the result
from the momentum fits. Clearly, there is a systematic difference at 
large times, which only decreases slowly as $y$ increases. 
The time propagators can be very well approximated  by a 
two-exponential fit, as shown in Fig. \ref{fig4}, but we do not have 
a deep understanding of the origin of the lighter states that seem to 
propagate at large times in this channel. 
What is clear from our data,
however, is that at small times, typically $t < b - 2 b$, the propagators
are very well reproduced by the hopping result and are thus dominated
by the propagation of a free massive fermion. 
In the future we will  explore the possibility that the deviations from the 
hopping result could be explained by the presence of fermion-scalar 
interactions. As explained in \cite{smit1} an effective 
Lagrangian for the neutral field compatible with the global symmetry is
\begin{eqnarray}
{\cal L}_{eff} \rightarrow \bar{\Psi}^{(n)}( \not\!\partial + m_n ) \Psi^{(n)}
+ \lambda \;\bar{\Psi}^{(n)}_L \gamma_\mu \Psi^{(n)}_L ( \omega \partial_\mu \omega^\dagger - \omega^\dagger \partial_\mu \omega ).
\label{neueff}
\end{eqnarray}
OC studies did not find evidence for such interactions. In fact, 
when the fermions are much lighter than the scalar fields, this  
is to be expected from FNN arguments, since the scalars should decouple from
the light spectrum, but when the fermions
are more massive than the scalars as in our case, there is no reason to believe that these interactions are absent. 

\begin{figure}
\begin{center}
\mbox{\epsfig{file=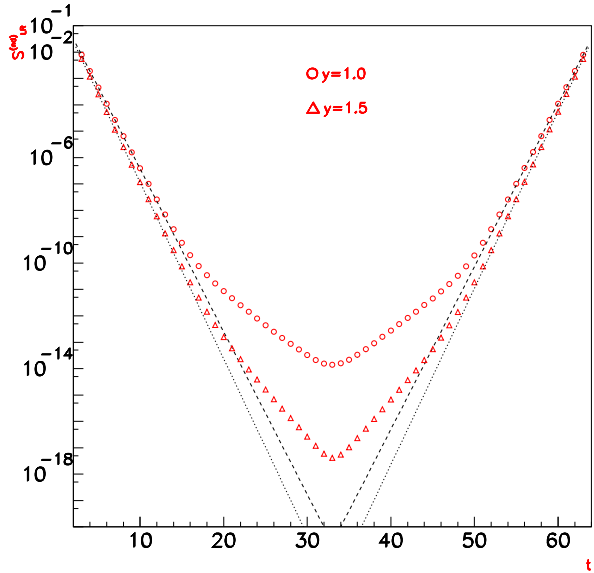,width=3.5in,height=3.5in}}
\end{center}
\caption[]{$S^{(nd)}_{LR}$ propagators in time for $y = 1, 1.5$ in a $8_8$ lattice. 
The dotted and dashed lines, which deviate at large times, are the results of the momentum fits of Table 2.}
\label{fig4b}
\end{figure}

The results for the neutral doubler are also shown in Table 2. As expected, 
there is good agreement with the hopping result in momentum space. The 
large-time behaviour of these
propagators, however, also shows an anomalous behaviour: the effective mass
at large times clearly decreases to a smaller value, see Fig. \ref{fig4b}. 
In the case of the doublers, a scalar-fermion interaction of the form
(\ref{neueff}) means that the doubler is not stable. It can decay 
into a light mode and two scalars, since it is heavier.   
This coupling is 
quite weak because in such a decay the scalars carry large
momenta and, as we have seen, the large Fourier modes of the scalar fields
are strongly suppressed. Empirically, the overlap with the lighter states is
very small:  $Z \sim 10^{-6}$. 

To summarize, we can conclude from our data that the first order 
hopping expansion provides a good approximation for the neutral propagators
at large momenta and small times, in the large-$y$ region.
At large times, the effective mass decreases and it is not clear whether 
there is an asymptotic {\it free} 
fermion in this channel. Notice that the FNN argument is not at work here, because 
the fermion mass is much larger than the scalar correlation length. In any 
case, the fact that the propagators are well reproduced by the 
hopping expansion at small times ($t < b - 2 b$) implies that,
by choosing $f/b$ small enough, the correlations can 
get negligibly small at $t \sim b$ and are not
expected to affect long-distance observables, if a light sector was 
present in the theory. 

\begin{figure}
\begin{center}
\mbox{\epsfig{file=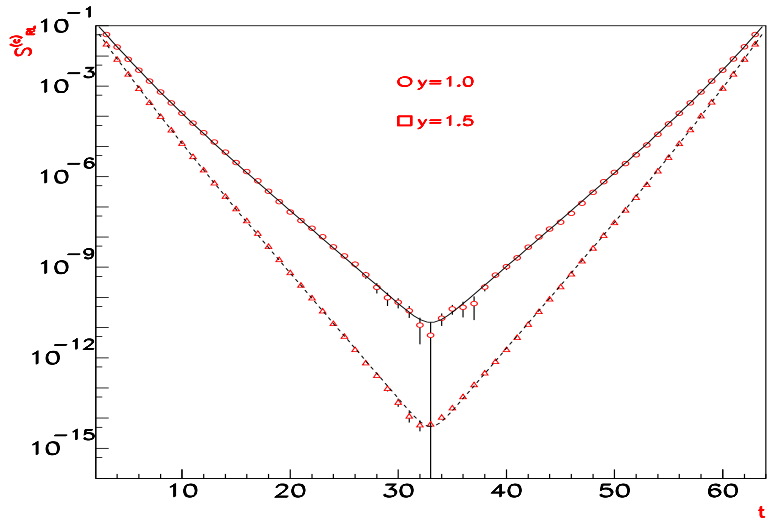,width=3.5in,height=3.5in,bbllx=60pt,bblly=-30pt,bburx=500pt,bbury=600pt}}
\end{center}
\caption[]{$S^{(c)}_{RL}$ propagators in time for $y = 1,1.5$. The lines
following the data points are the result of the two-exponential fits.}
\label{fig5}
\end{figure}

The charged propagators, both doubler and non-doublers, also behave as massive Dirac fermions. However, in this
case the hopping approximation is not a good approximation 
even at small times. This is similar to what was found in OC studies. 
In Fig. \ref{fig5} we show the RL 
components of the charged propagator, together with a two-exponential fit. 
\begin{table}
\begin{center}
\begin{tabular}{|c|c|c|}\hline
y & $m^{(n)}$ ($f$-units) & $m^{(c)}$ ($f$-units) \\ \hline
1. & 0.538(4) & 0.745(27) \\
\hline
1.2 & 0.653(7) & 0.868(43)\\
\hline
1.5 & 0.816(3) & 0.982(21) \\
\hline
\end{tabular}
\caption[]{Effective masses from a single-exponential fit at large times 
of the neutral and charged $LR$ propagators for large $y$, for a lattice size 
of
$8_8$.}
\end{center}
\end{table}
In Table 3, we compare the effective masses obtained from a single-exponential
fit in the large-time region of the neutral and charged $LR$ propagators.
The difference between them is of the order of the scalar mass from 
Table 1. This is consistent with the charged propagator being a two-particle state, $\omega^\dagger \Psi^{(n)}$, as favoured by OC studies.  

\subsubsection{Small-$y$ Regime}

As we decrease $y \leq 1$, the different chiral components of both the 
light neutral and charged propagators start to differ, signalling a
violation of parity. Two components start getting lighter than the others: $S^{(n)}_{RL} = S^{(p)}_{RL}$ and $S^{(c)}_{LR} = S^{(p)}_{LR}$ (i.e. 
the expected physical fields). 
From our data we cannot establish a transition line in which parity 
violation shows up. It seems that the behaviour of the propagators changes
in a smooth way, the violations of parity being present also at large $y$ and
getting larger smoothly as $y$ is decreased. 

There is, however, a critical value of $y_c$ below which the two lighter
components $S^{(n)}_{RL}$ and $S^{(c)}_{LR}$ get actually massless at
finite lattice spacing. This is the onset of the announced chiral phase.  
It is not clear what determines the value of $y_c$. 
There is one obvious candidate, which is the boson cutoff, i.e. $f/b$ in $f$-lattice units. Although our numerical results are not 
definite on this point, they seem to indicate that $y_c$ is related
to this scale. The physical picture behind this expectation is 
nothing but the FNN conjecture that at distances 
larger than $b/f$ in $f$-units, the scalars should decouple. The 
correlation length of the light fermions in the strong phase is $\sim y$, so 
when $y < f/b$ the scalars decouple from the light fermions and a chiral
phase appears.

\begin{figure}
\begin{center}
\mbox{\epsfig{file=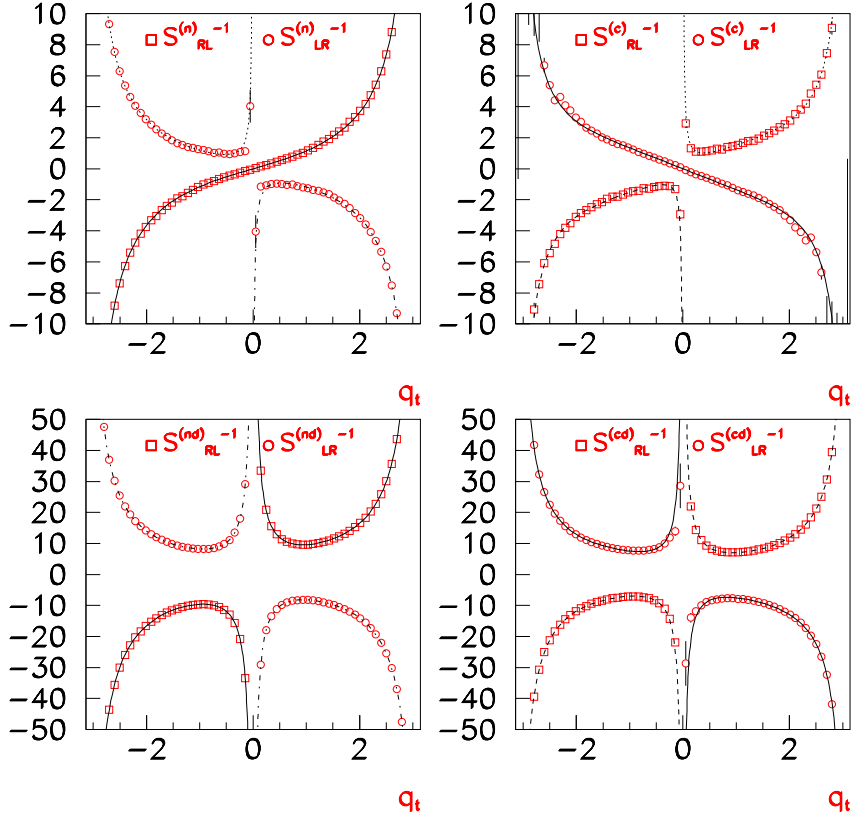,width=5in,height=5in}}
\end{center}
\caption[]{Inverse propagators $(S^{(n),(nd)})^{-1}$ and $(S^{(c),(cd)})^{-1}$ in momentum space in an 
$8_8$ lattice for $y=0.05$.}
\label{fig6}
\end{figure}

 As explained in the previous section, in order to study this phase, it
is better to go to momentum space, where the limits $q \rightarrow 0,\pi$ of
the propagator are known. In Fig. \ref{fig6}, we present the RL and LR components of the neutral and charged 
inverse propagators for $y=0.05$. Clearly the $S^{(n)}_{RL} = S^{(p)}_{RL}$ 
and $S^{(c)}_{LR} = S^{(p)}_{LR}$ propagators have
poles only at $q \rightarrow 0$, as expected from massless undoubled fermions.
On the other hand, the components $S^{(n)}_{LR}$ and $S^{(c)}_{RL}$ show no
poles, in agreement with them being two particle states composed of a massive
scalar particle and a massless fermion (\ref{chiexp}). 

\begin{figure}
\begin{center}
\mbox{\epsfig{file=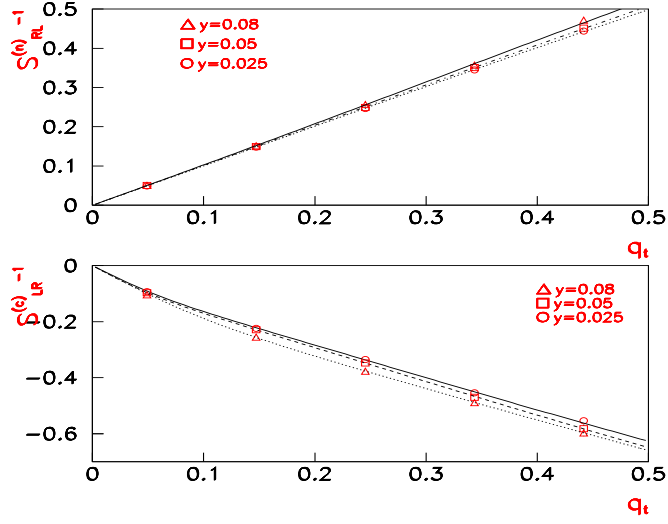,width=3in,height=3.5in,bbllx=60pt,bblly=-30pt,bburx=500pt,bbury=600pt}}
\end{center}
\caption[]{$(S^{(n)}_{RL})^{-1}$ and $(S^{(c)}_{LR})^{-1}$ at small momentum for several 
values of $y$ in the chiral phase.}
\label{fig7}
\end{figure}

In Fig. \ref{fig7}, we present a zoom of the massless components in the small 
momentum region for several
values of $y$. The continuum curves are fits of the form (the spatial momentum
being zero),
\begin{eqnarray}
S^{(n,c)}_{RL,LR}(q) = \frac{Z_{R,L}}{\sin(q)} + Z'_{R,L} \frac{\sin(q)}{\sin(q)^2+{m'}_{R,L}^2}.  
\label{chifit}
\end{eqnarray}
The second term corresponds to higher states, which should decouple 
in the continuum limit. Their masses are expected to be of the order of 
the boson cutoff or larger.  It is clear from the figures that the 
contamination of the charged chiral mode is larger than that 
of the neutral one.  
We show the results of these fits for several lattice
sizes in Table 4.
\begin{table}
\begin{center}
\begin{tabular}{|c|c|c|c|c|}\hline
Size & $Z_R$ & $Z_L$ & $m'_L$ ($b$-units) & $\frac{Z'_L}{Z_L}$ \\ \hline
$6_6$& 0.994(1) & 0.543(12) & 1.16(17) & 0.384(25) \\
$6_8$ & 0.994(1) & 0.668(17)& 1.27(34) & 0.257(23) \\
$8_4$ & 0.994(3) & 0.168(33)& 0.95(22) & 2.245(54) \\ 
$8_8$ & 0.996(1) & 0.494(21)  & 1.08(13) & 0.603(84)\\
$10_8$ & 0.995(1) & 0.485(40) & 1.22(44) & 0.52(12) \\
\hline
\end{tabular}
\caption[]{$Z_{L,R}$, $Z'_L/Z_L$ and $m'_L$ for various lattice sizes. $y$ is
kept fixed to $0.2$ in units of $b$.}
\end{center}
\end{table}
The volume effects as  $L/b$ changes, for fixed $f/b$, are quite small and
 it seems clear that the 
contaminating mass $m'$ does not decrease as the volume is increased.
On the other hand, for fixed $L/b$, the contamination increases considerably
for the charged propagator, when the cutoff ratio is increased.

\begin{figure}
\begin{center}
\mbox{\epsfig{file=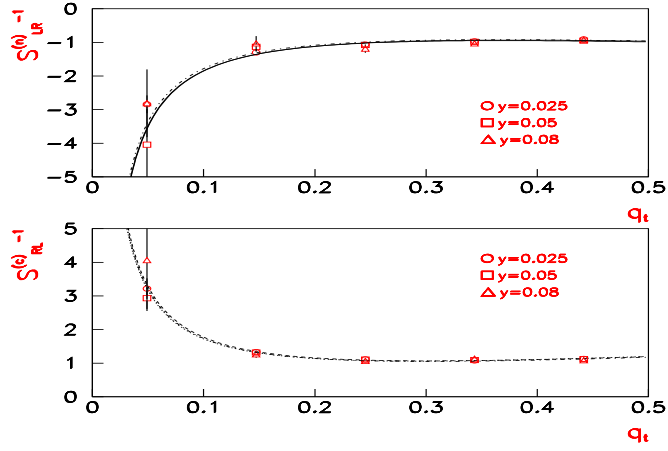,width=3in,height=3in,bbllx=60pt,bblly=-30pt,bburx=500pt,bbury=600pt}}
\end{center}
\caption[]{$(S^{(n)}_{LR})^{-1}$ and $(S^{(c)}_{RL})^{-1}$ at small momentum for several 
values of $y$ in the chiral phase. The lines are fits to (\ref{chifit2}), which 
give $m^{(n)}_L= 2.46(13), 2.46(35), 2.10(55)$ and $m^{(c)}_R= 2.32(10), 2.19(21), 2.07(29).$, in increasing order of $y$.}
\label{fig7b}
\end{figure}

In Fig. \ref{fig7b}, we zoom the low-momentum behaviour of $S^{(n)}_{LR}$ and
$S^{(c)}_{RL}$. It is important to make sure that these propagators 
do not develop poles as $y$ gets smaller or as 
the continuum limit is approached, rendering the theory vector-like.
 It is not clear how to fit these propagators at 
zero momentum, since they are expected to be multiparticle states. 
We have used the simplest ansatz, 
\begin{eqnarray}
S^{(n,c)}_{LR,RL}(q) = \tilde{Z}_{L,R} \frac{\sin(q)}{\sin(q)^2 + (m^{(n,c)}_{L,R})^2}, 
\label{chifit2}
\end{eqnarray}
which empirically 
gives a good fit near $q = 0$. In Table 5, we present the masses
obtained from these fits for various lattice sizes.
\begin{table}
\begin{center}
\begin{tabular}{|c|c|c|}\hline
Size & $m^{(n)}_L$ ($b$ units) & $m^{(c)}_R$ ($b$ units)   \\ \hline
$6_6$& 2.26(11) & 2.08(78)  \\
$6_8$ & 2.19(14) & 2.03(11) \\
$8_8$ & 2.46(13) & 2.32(10) \\
$10_8$ & 2.59(51) & 2.03(19) \\
\hline
\end{tabular}
\caption[]{$m^{(c)}_R$ and $m^{(n)}_L$ from a fit to (\ref{chifit2}) for various lattice sizes, in units of the boson cutoff $b^{-1}$; $y$ is kept fixed
to $0.2$ in $b$ units.}
\end{center}
\end{table}
It is clear that these masses are of the order of the boson cutoff, as
expected from (\ref{chiexp}). Also, there is no sign of the masses getting
smaller as the volume increases or as $y$ is decreased.

For the neutral 
doublers, we find that the large-momentum ($q > \pi/b$) and small-time ($t < b$) behaviour corresponds to that of a free massive fermion,
which supports the picture that, due to the strong coupling between the 
doubler modes and the scalars, the neutral composite fermion (\ref{composites}) is formed. 
In Table 6, we present the fits of the neutral doubler inverse propagators
to the free Wilson formulae for large momentum ($q > \pi/b$). The
effective $r$ is in perfect agreement with the hopping result. 
The effective $y$ on the other hand differs (especially for the 
RL component). This could be due
to a low-momentum bias, since the hopping result
is not a good approximation for momenta of the order of the 
boson cutoff.
\begin{table}
\begin{center}
\begin{tabular}{|c|c|c|c|c|c|c|c|}\hline
y & Prop & & RR & RL & LR & LL & Hopp.\\ \hline
0.025   & $S^{(nd)}$ & Z &1.083(6)& 0.992(16) & 1.173(8)& 1.087(3) & \\  
    &           & $y_{eff}$ &0.033(5)& 0.004(4) & 0.030(4) & 0.033(1) & 0.027\\
    &           & $r_{eff}$ &1.082(5)& 1.090(8) & 1.082(4) & 1.086(3) & 1.085\\
\hline
0.05 & $S^{(nd)}$ & Z &1.084(5)& 0.993(16) & 1.173(8)& 1.087(2)& \\  
    &           & $y_{eff}$ &0.061(3)& 0.024(4) & 0.057(1) & 0.059(5)& 0.054\\
    &           & $r_{eff}$ &1.082(4)& 1.090(8) & 1.082(4) & 1.086(3) & 1.085\\
\hline
0.08   & $S^{(nd)}$ & Z &1.082(4)& 0.997(17) & 1.172(8)& 1.087(4) & \\  
    &           & $y_{eff}$ &0.092(1)& 0.056(5) & 0.089(1) & 0.092(1) & 0.087\\
    &           & $r_{eff}$ &1.081(4)& 1.092(8) & 1.082(4) & 1.086(4) & 1.085\\ 
\hline
\end{tabular}
\caption[]{Fits of the neutral doubler propagators in momentum space ($q > \pi/b$) to the free Wilson formula for small $y$ in an $8_8$ lattice.}
\end{center}
\end{table}

At large times or small momenta, there is evidence of 
contamination from lighter states, which are probably multiparticle states.
This is not so surprising since the neutral composite is not expected to 
be stable in this phase.
Possible effective interactions, compatible with the symmetries that 
connect the light and doubler sectors are given by
\begin{eqnarray}
%g_c \;\{\; \bar{\Psi}^{(cd)} \omega^\dagger P_R\; \partial_{\mu}^2 \Psi^{(p)} % +
%\bar{\Psi}^{(p)} \partial_{\mu}^2 (\;\omega P_L \Psi^{(cd)} ) \}\nonumber\\
g_n \;\{\; \bar{\Psi}^{(p)} \omega^\dagger P_R \;\partial_{\mu}^2 \Psi^{(nd)}  +
\bar{\Psi}^{(nd)} \partial_{\mu}^2 (\omega P_L \Psi^{(p)} ) \;\}.
\label{inter}
\end{eqnarray}
In the presence of these interactions,
 the neutral doubler channel $S^{(nd)}_{RL}$ at large
times will be dominated by two-particles states. 
The decay of the neutral doublers into lighter particles however involves high momentum modes
of the scalar field, which are suppressed for TC, so we expect that 
they are relatively long-lived and thus dominate the propagators
at small times.  Notice that in the OC case, this does not have to be the 
case. 

\begin{figure}
\begin{center}
\mbox{\epsfig{file=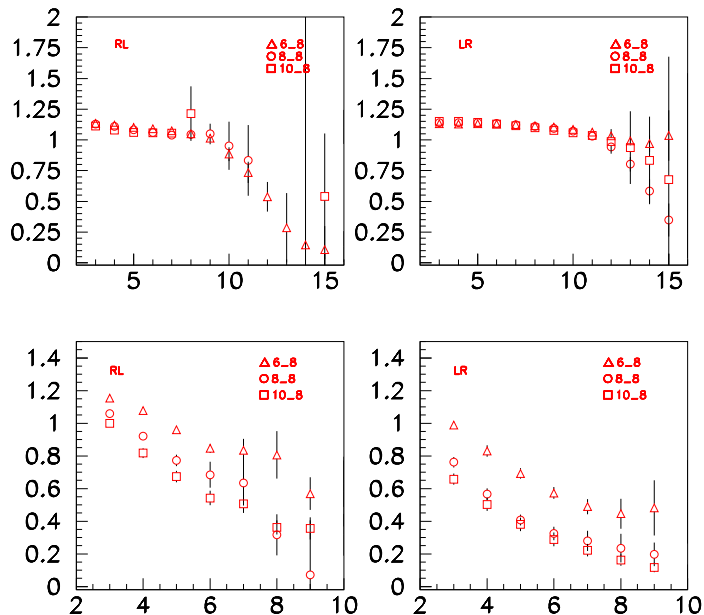,width=4in,height=4in}}
\end{center}
\caption[]{Effective mass of the neutral (up) and charged (down) doublers
for several lattice sizes and $y=0.025$.}
\label{fig17}
\end{figure}

In Fig. \ref{fig17} we
present the effective masses in the $RL$ and $LR$ neutral and charged
doubler channels for $t < b - 2 b$, for three values of $L/b$ and fixed $f/b$. The neutral effective mass  
has a nice plateau at small times. This is consistent with the assumption 
that this channel is dominated by the neutral composite of (\ref{composites}). On the other hand, the 
charged doubler does not show a plateau. The effective mass is of the order of the 
fermion cutoff at small times, but it is nowhere constant. This indicates
that the charged doubler composite is not formed and multiparticle
states are dominating this channel. 

Although our statistics is not good enough to attempt a quantitative 
description of the doubler propagators at large times, the fact 
that, at small times $t < b - 2 b$, the propagators decay as particles with 
masses of the order of the fermion cutoff implies that, after 
a few $f$-time slices, the correlation in the doubler channels is too 
small to influence the long-distance physics.
In the simplified 
model we are considering here there is no more long-distance physics
than the free propagation of massless fermions. 
In Figs. \ref{fig11} and \ref{fig11b}, we compare the decay in time of the 
various propagators,
neutral and charged respectively, for several lattice sizes. The massless
components have constant correlation as expected. 
All the remaining
propagators are not simple at large times, but they are clearly 
split from the light sector. For $f/b =1/8$, they are suppressed by at least 2--3 orders of magnitude. We 
notice that the effect of increasing $f/b$ is quite dramatic 
for the charged propagator (see last figure in Fig. \ref{fig11b}). In general
the splitting between the doubler and non-doubler sectors is larger for the
neutral channel. 

To  summarize, we have found that at small enough $y$ there is a chiral 
phase, in which the light spectrum consists of two massless fermions with 
chiral quantum numbers. Their propagators in momentum space show only one pole
at $q \rightarrow 0$, as expected from undoubled massless particles. The 
neutral doubler channel seems to be dominated by a massive fermion
 at small
times, while at large times it is probably dominated by multiparticle
states. The charged doubler  channel seems to be everywhere dominated by multiparticle
states. These states contain scalars, whose correlation
length is of the order of the boson cutoff $b^{-1}$; thus 
they are expected to decouple in the continuum limit. In order to show that this is indeed
the case, it will be necessary to consider the case where gauge interactions
are switched on, since only in this case is there a physical mass gap that
can be used to define the continuum limit of the lattice model.  
We have not found important volume effects as 
$L/b$ is increased, although there is a clear dependence on $f/b$: the 
splitting between the light sector and the heavy one becomes smaller
as $f/b$ is increased. This is a clear indication of the importance of the
cutoff separation.  

\begin{figure}
\begin{center}
\mbox{\epsfig{file=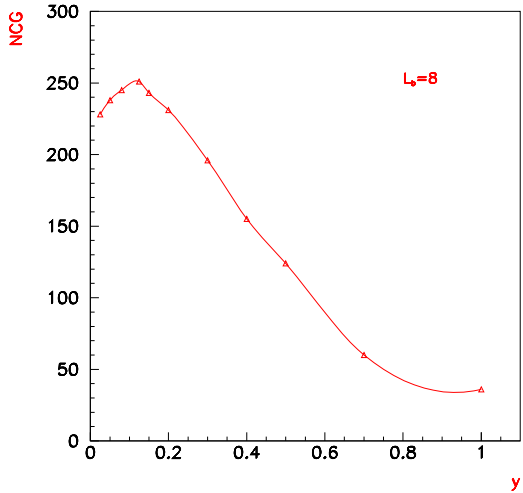,width=3in,height=3in}}
\end{center}
\caption[]{Number of conjugate-gradient iterations in the inversion of the 
neutral propagator as a function of $y$ in an $8_8$ lattice.}
\label{fig8}
\end{figure}

\begin{figure}
\begin{center}
\mbox{\epsfig{file=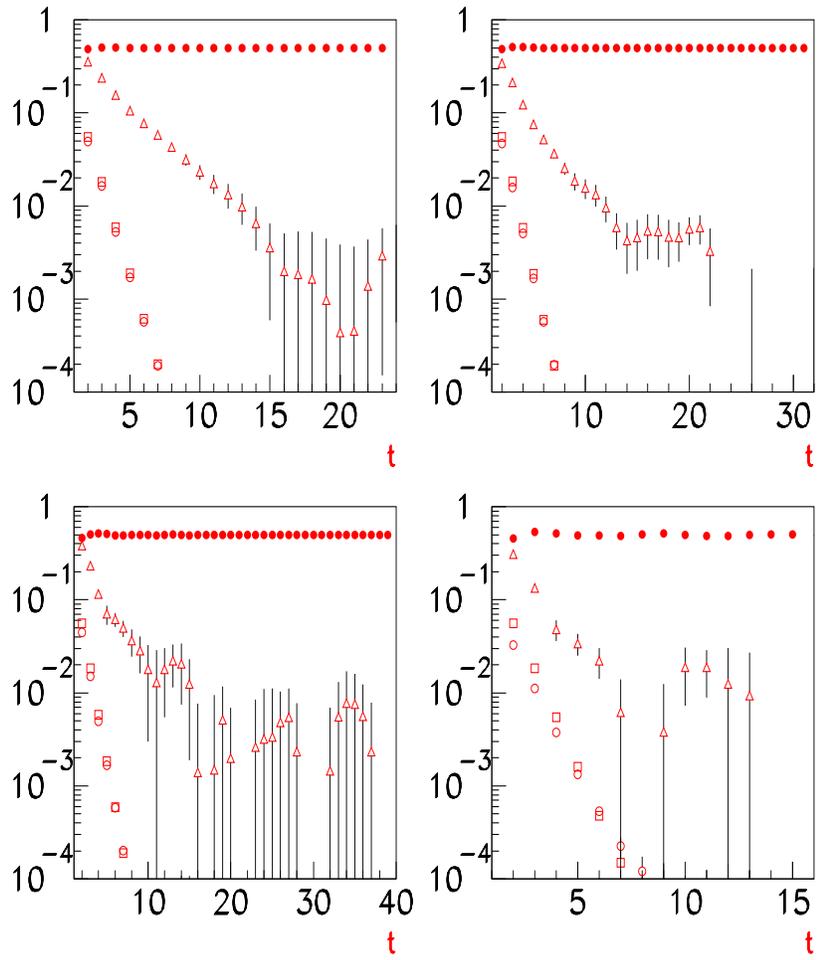,width=5in,height=6in}}
\end{center}
\caption[]{$S^{(n)}_{RL}$ (full circles), $S^{(n)}_{LR}$ (triangles), 
$S^{(nd)}_{RL}$ (circles), $S^{(nd)}_{LR}$ (squares) for several lattice
sizes: $6_8$, $8_8$, $10_8$ and $8_4$, for $y=0.025$.}
\label{fig11}
\end{figure}

\begin{figure}
\begin{center}
\mbox{\epsfig{file=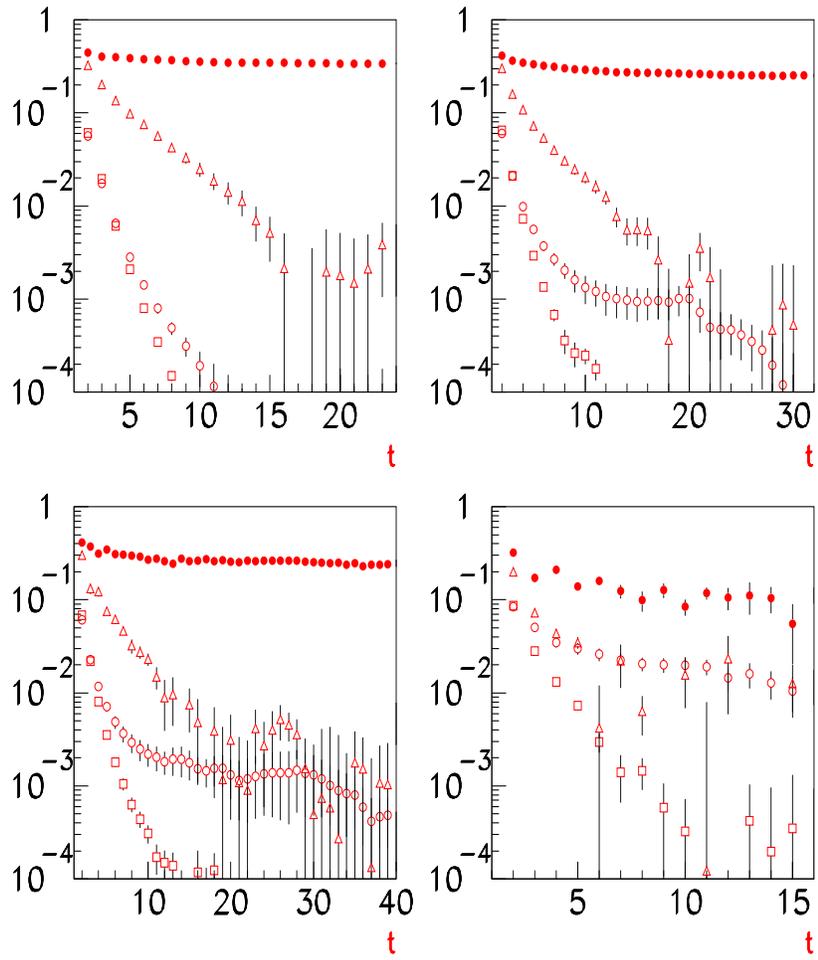,width=5in,height=6in}}
\end{center}
\caption[]{$S^{(c)}_{LR}$ (full circles), $S^{(c)}_{RL}$ (triangles), 
$S^{(cd)}_{LR}$ (circles), $S^{(cd)}_{RL}$ (squares) for several lattice
sizes: $6_8$, $8_8$, $10_8$ and $8_4$, for $y=0.025$.}
\label{fig11b}
\end{figure}

We have not analysed in detail the intermediate region of couplings
$y_c < y < O(1)$, and in particular, we have not addressed the important
issue of the nature of the transition at $y_c$. 
More statistics will be needed for this. However, we have monitored 
one observable that has proved useful in the location of  crossover or 
transition lines in previous studies of Wilson-Yukawa
models \cite{wy}. This is simply
the number of conjugate-gradient iterations (NCG) needed in 
the propagator inversion for a fixed precision. The naive expectation is  
that the required number of iterations
should increase as $y$ decreases, because the lowest eigenvalue of the fermion matrix decreases with $y$. This changes however when we enter the chiral phase, since the
light mode is massless and the lowest eigenvalue
is fixed by the IR cutoff, which is the lowest lattice momentum (remember
we use AP boundary conditions in the time direction). Thus 
the number of iterations should not keep increasing for $y < y_c$. In Fig. \ref{fig8}, 
we show the NCG 
for different values of $y$. There seems to be a maximum very close to 
the value where we believe $y_c$ should be, according to the behaviour of the 
propagators. 
The position of the maximum is in fact near $y = 0.15$ in $f$ units, which is
not far from $f/b = 1/8$, as naively expected. 

\section{Effects of unquenching}

It is clear that unquenching in the OC case can have dramatic effects. By 
naive power counting, fermion loops can induce $O(1)$ 
corrections to the potential of the $\omega$ fields. In order to be in the
FNN regime, tunings of $\kappa$ and probably other relevant operators 
of dimension 4 that are generated at higher orders  might be 
needed.

As explained in \cite{us1}, in the TC case the situation is very different.
Due to the suppression of the high-momentum modes of the scalar fields, 
it can be shown that the corrections induced by fermion loops on the scalar 
potential are down by powers of the two-cutoff ratio\footnote{As explained in \cite{us1} the fermion determinant is regulated in a way in which the 
breaking of gauge invariance is restricted to the phase of the determinant. The absolute value can be defined in a gauge-invariant way, otherwise 
one-loop subtractions would be needed.}, $O(f/b)^2$. For this 
reason it is expected that unquenching is not going to change the phase
diagram of the fermion-scalar model in any important way.
 A way to 
quantify this is by looking at how the phase of the fermion determinant for the 
action (\ref{u1}) changes under a gauge transformation. The magnitude of this
change is a measure of the strength of 
the couplings of the scalar fields $\omega$, induced by the fermion loops. At this stage, however, we have to enlarge the 
fermion content of the model we are considering in order to cancel 
gauge anomalies. Otherwise the changes in the phase are of $O(1)$ and 
the two-cutoff method is of no use, nor any other, of course. A simple 
choice is the $11112$ model, with four left-handed fermions with $U(1)$ charge
$q_L = 1$ and one right-handed fermion with charge $q_R = 2$. Gauge anomalies cancel,
because
\begin{eqnarray}
\sum_i q^2_{i,L} = \sum_i q^2_{i,R}.
\end{eqnarray}

\begin{figure}
\begin{center}
\mbox{\epsfig{file=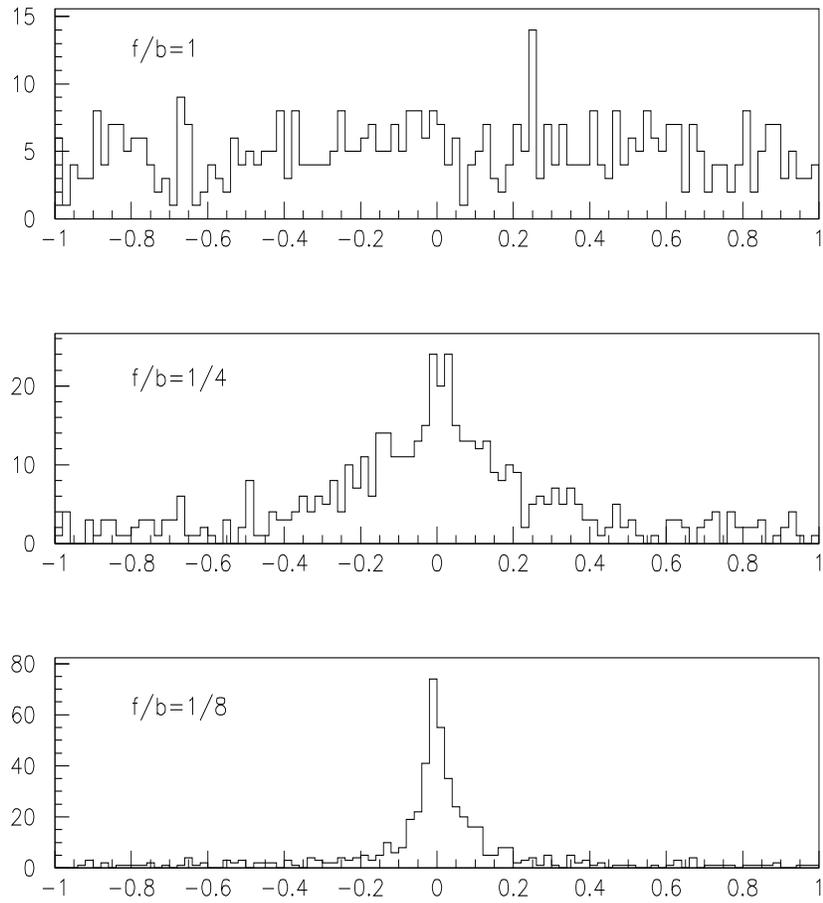,width=5in,height=6in}}
\end{center}
\caption[]{Change in the phase of the fermion determinant under random 
gauge transformations in lattice sizes $4_1$, $4_4$ and $4_8$.}
\label{fig13}
\end{figure}

In Fig. \ref{fig13}, we present a histogram of the change in the phase of the  fermion determinant in a lattice of $L/b = 4$ and several values of 
$f/b = 1, 1/4, 1/8$, under a random $b$-lattice gauge transformation, starting
from the trivial configuration. As expected, the change in the phase of the fermion
determinant gets smaller with the ratio $f/b$. For arbitrarily small $f/b$, it
is clear that the FNN conditions are still satisfied in the full theory. A
full unquenched simulation would be needed to determine how small
this ratio has to be in practice. 

Fermion loops are also expected to induce renormalization of the 
Yukawa couplings $y$ and $r$ at higher orders. For the same reason as before, it can be shown \cite{us1}
that the corrections to these couplings are down by powers of $(f/b)^2$. 
Thus, for small enough $f/b$, no tunning of the bare Yukawa couplings is 
needed to remain in the chiral phase.  

\section{Conclusions} 

We have presented numerical evidence in favour of the existence of a truly 
chiral phase in a two-cutoff formulation of a quenched 
2D Wilson-Yukawa model with 
a global $U(1)_L \times U(1)_R$. The Yukawa phase diagram is 
different from the one found in the one-cutoff formulation of this model in 
that there is a new phase for $r = O(1)$ and $y \ll 1$, where doublers
are split from the light spectrum, which is composed of
 massless fermions with 
chiral quantum numbers. When one of the chiral symmetries is gauged, 
the continuum limit of this lattice model is expected to give rise to a
 chiral gauge theory. 

The main new feature of the two-cutoff formulation is that 
the high Fourier modes of the scalar fields are naturally cut off at a scale smaller than the momenta of the fermion doublers. As a consequence the 
Wilson-Yukawa term is truly a perturbative coupling for the ligh-fermion 
mode ($p \rightarrow 0$), while it is effectively a large $O(r)$ 
Yukawa coupling for the doubler modes ($p \rightarrow \pi$). Near 
$p \rightarrow 0$, the chiral fermion propagators are well reproduced by weak Yukawa perturbation
theory and have poles at $p \rightarrow 0$, while at large momenta they are
better described by the strong Yukawa expansion, which predicts no doubler 
poles (since all fermions are massive in this approximation).

All our results have been obtained in the symmetric or VX phase (PM in 4D) of the 
scalar dynamics. This is the relevant phase for describing a chiral gauge
theory. In the case of a spontaneously broken gauge theory in 4D, the relevant phase would be a ferromagnetic (FM) one. If a FM counterpart of the chiral 
phase found here exists,
it would lead to a successful discretization of the Standard Model. This 
interesting case will be considered elsewhere.

{\bf Acknowledgements}
We thank J.L. Alonso, M.B. Gavela, M. Guagnelli, K. Jansen, R. Sundrum, M. Testa and T. Vladikas for 
very helpful discussions/suggestions. 
LPTHE is laboratoire associ\'e au CNRS.
\vspace{1cm}

\setcounter{section}{0}
\setcounter{equation}{0}
\renewcommand{\thesection}{Appendix \Alph{section}.}
\renewcommand{\theequation}{A. \arabic{equation}}

\section{}

We present here the explicit formulae for the compact $U(1)$ interpolation in 
2D. For details on how to construct this interpolation the reader is 
referred to \cite{us2}. The methods are similar to those used in 
\cite{luscher1}.

The $b$-lattice sites are denoted by $s$. For an arbitrary
 gauge  configuration 
on the $b$-lattice, 
we can uniquely write each lattice link variable in the form
\begin{equation}
U_{\mu}(s) = e^{i A_{\mu}(s)}, ~~|A_{\mu}(s)| < \pi, 
\end{equation}
where we are neglecting the measure-zero set of lattice fields, where at
least one of the link variables equals exactly $-1$. This assignment
defines an obvious logarithm function for group elements different from 
$-1$: 
\begin{equation}
A_{\mu} = -i \log U_{\mu},  ~~|A_{\mu}| < \pi.
\end{equation}
The link variables on the $f$-lattice can also be writen as 
\begin{equation}
u_{\mu} = e^{i a_{\mu}},  ~~|a_{\mu}| < \pi.
\end{equation}

Defining the $f$-lattice sites within each $b$-lattice plaquette 
as $x = s + t_1 \hat{1} + t_2 \hat{2}$, where $0 \leq t_1, t_2 \leq 1$, the
final result is,  
\begin{eqnarray}
 a_1 &=&\frac{f}{b}\;\{\; (1 -\; t_2)\; ( A_1(s) - 2 \pi N_2(s) ) + t_2
\; ( A_1(s+\hat{2}) - 2 \pi N_2(s+\hat{2})) \;\}\nonumber\\
a_2 &=& \frac{f}{b}\;\{\;(1 -\; t_1) \;( A_2(s) + 2 \pi N_1(s) )
+ t_1 \; ( A_2(s+\hat{1}) + 2 \pi N_1(s+\hat{1}) ) \;\} 
\label{interqed}
\end{eqnarray}
Notice that the fields at the boundary of the plaquette satisfy $a_{\mu} = 
f/b \;A_{\mu}$.

Except for the contribution of the $N_{\mu}$ fields, this is simply a
linear interpolation, as naively expected. 
The $N_{\mu}$ fields are integers defined by the following equation:
\begin{eqnarray}
\sum_{\mu} \;\;N_{\mu}(s+\hat{\mu}) - N_{\mu}(s) = N(s) 
\label{div}
\end{eqnarray}
where $N(s)$ are the winding numbers of the $a_{\mu}$ field 
defined on the boundary of the plaquette (to be more precise the winding
of the $f\rightarrow 0$ limit of $a_{\mu}$ on the boundary), i.e.
\begin{eqnarray}
I[ \frac{A_2(s) + A_1(s+\hat{2}) - A_2(s+\hat{1})- A_1(s)}{2 \pi} ]
\label{wdef}
\end{eqnarray}
with $I[y] \equiv$ nearest integer to $y$.
 The existence of these fields
is of course related to the compactness of the group. In particular, the 
geometrical definition of topological charge is simply given by, 
$\sum_s N(s)$. In this paper we are only concerned with topologically trivial
configurations since $g = 0$. 
Even in the case of trivial topology, a non-measure zero set of  
lattice configurations will have $N(s) \neq 0$. The number of non-zero 
windings grows with the $b$-lattice volume. 

\begin{figure} 
\begin{center}
\mbox{
\epsfig{file=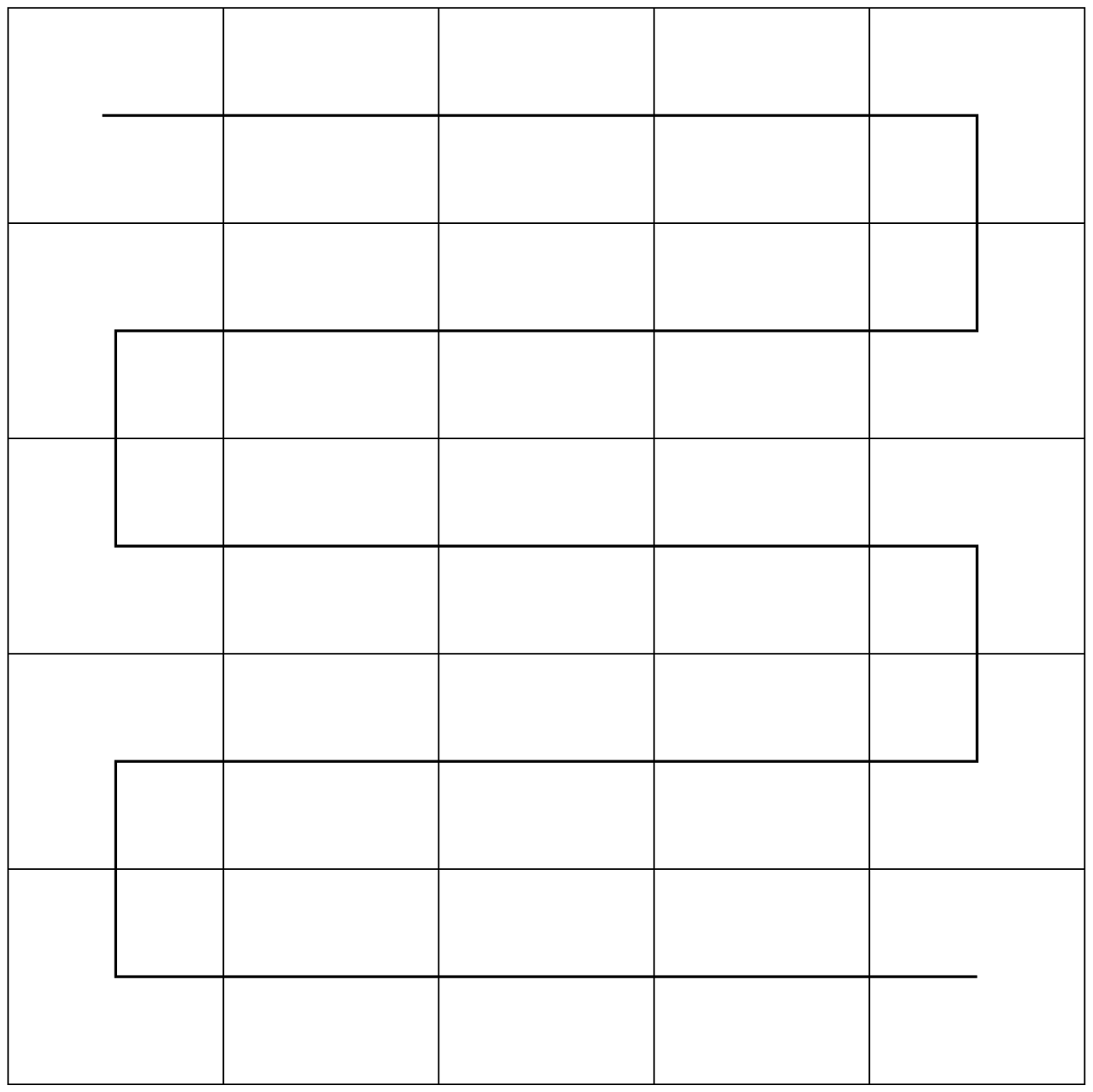,height=4cm,width=4cm}}
\end{center}
 \caption[]{``Gauge'' choice used in solving (\ref{div})
for compact $QED_2$. The $N_{\mu}$ corresponding to links
 that are not intersected by the path are zero.}
 \label{fig:lattice}
\end{figure}

Equation (\ref{div})
has a unique solution in the ``gauge'' depicted in Fig. \ref{fig:lattice}.  
All the links that are not crossed by the path have the associated
$N_{\mu}(s)$ set to zero (remember that $\hat{\mu}$ is $orthogonal$ to the
link in our notation). In order to ensure the discrete 
lattice symmetries, however, 
we have to also consider the interpolations obtained in the maximal 
gauges that are obtained from that in Fig. 1 by lattice symmetry transformations (i.e. $90^\circ$ rotations and temporal and spatial inversions). There 
are eight of these maximal trees in total. Ideally, the maximal path could be
chosen randomly for each $b$-lattice configuration. 
This is however more complicated to implement, so we have used the 
first method.

We also need the interpolation of $b$-lattice gauge transformation, defined as
\begin{eqnarray}
u[U^\Omega](x) = u^\omega[U](x),
\end{eqnarray}
but in the simpler global limit, i.e.
\begin{eqnarray}
u[I^\Omega](x) = u^\omega[I](x) = I^\omega(x), 
\end{eqnarray}
where $I$ is the identity configuration, i.e. $U_{\mu}(s) = 1$ or $u_{\mu}(x)=1$.
For the prescription (\ref{interqed}), there is a unique solution for $\omega$. Defining
\begin{eqnarray}
\Omega(s) \equiv e^{i \Phi(s)}, \;\;\;\;\; \omega(x) \equiv e^{i \phi(t_1,t_2)},
\label{omdef}
\end{eqnarray}
the solution is, 
\begin{eqnarray}
\phi(t_1,t_2) = \Phi(s) + (A_1(s) - 2 \pi N_2(s)) \;t_1 + ( A_2(s) + 2 \pi N_1(s) )
\;t_2 \nonumber\\
+ (A_2(s+\hat{1}) + 2 \pi N_1(s+\hat{1}) - A_2(s) - 2 \pi N_1(s)) \; t_1 t_2  
\label{interom1}
\end{eqnarray} 
with 
\begin{equation}
A_{\mu} \equiv -i \log (\Omega^\dagger(s) \Omega(s+\hat{\mu})),  ~~|A_{\mu}| < \pi,
\label{pureg}
\end{equation}
and the $N_{\mu}$ fields are defined by (\ref{div}), (\ref{wdef}) and (\ref{pureg}).

It is easy to see that $\phi(0,0) = \Phi(s)$ and $\omega$ is thus 
a smooth interpolation of the gauge transformation $\Omega$.

\end{document}